\documentclass[eqsecnum,floats,aps,prd,floatfix,titlepage,tightenlines]{revtex4} 

\usepackage{graphicx}
\usepackage{graphics}
\usepackage{bm}
\usepackage{amssymb}
\usepackage{amsmath}
\usepackage{mathrsfs}

\begin{document}
\title{Cosmological Particle Production: A Review}
\author{L. H. Ford}
\email{ford@cosmos.phy.tufts.edu}
\affiliation{Institute of Cosmology, Department of Physics and Astronomy\\
Tufts University, Medford, MA 02155, USA
\vskip 0.5in}
\date{\today}

\begin{abstract}
This article will review quantum particle creation in expanding universes. The emphasis will be on the basic physical
principles and on selected applications to cosmological models. The needed formalism of quantum field theory in
curved spacetime will be summarized, and applied to the example of scalar particle creation in a spatially flat universe.
Estimates for the creation rate will be given and applied to inflationary cosmology models. Analog models which
illustrate the same physical principles and may be experimentally realizable are also discussed. 
\end{abstract}

\maketitle

\section{Introduction and Basic Concepts}
\label{sec:basic}

\subsection{Preliminary Remarks}
\label{sec:prelim}

This review will cover selected aspects of quantum field theory in curved spacetime, with emphasis
on quantum particle creation in expanding universes. For broader treatments of quantum field theory 
in curved spacetime, see, for example, the books by Birrell and Davies~\cite{BD} and by Parker 
and Toms~\cite{PT}. As much as possible, the focus will be on the key physical ideas more than
the mathematical formalism, but a certain amount of the latter is essential. I have attempted to give
a balanced selection of references, but it is not possible to cite all of the numerous papers which have 
been written on cosmological particle creation and the related topics covered in this review.
I apologize in advance to the authors whose work has not been cited.

\subsection{An Overview}
\label{sec:overview}

This review will deal with quantized fields propagating on a classical background spacetime other than Minkowski
space. As general relativity describes a gravitational field as a curved four dimensional spacetime, this is a means
to describe the interaction of quantum particles with a background  gravitational field. One of the key features of
quantum field theory in Minkowski spacetime is heavy reliance on Poincare symmetry, which is broken by the
background  gravitational field. If the gravitational field is spatially inhomogeneous, then space translation symmetry
is broken, and linear momentum in no longer conserved. The initial and final momenta of particles scattering from
a localized gravitational field need not be equal, as the gravitational field can absorb some momentum, as in the case 
of a particle scattering from a black hole.  

If the background gravitational field is time-dependent, as in the case of an expanding universe, then time translation
symmetry is broken. Now the energy of the quantum particles need not be conserved, and the quantum field may
absorb energy from the background. This will play a key role in cosmological particle creation.

The outline of this review is as follows: The formalism of field quantization in a curved spacetime is reviewed in
Sec.~\ref{sec:quantization}, and the mathematical tools needed to describe quantum particle creation will be developed
in Sec.~\ref{sec:particle-creation}. This will include Bogolubov transformations, and a discussion of some exactly
soluble models, as well as of a perturbative approximation method. Some estimates of particle creation rate will be
discussed. Section~\ref{sec:state-entropy} will treat the quantum state of the created particles, and discuss decoherence
and entropy production. The implications of quantum particle creation for inflationary cosmology, and for subsequent epochs
of the universe, are discussed in Sec.~\ref{sec:cosmology}. The backreacion of the created particle on the expansion of
the universe is the topic of Sec.~\ref{sec:backreact}. Section~\ref{sec:analog} discusses some other physical systems
which rely upon the same principles as cosmological particle creation, and may be able to be studied in the laboratory.
Section~\ref{sec:final} summarizes the key topics covered in the review.

\subsection{Quantization in Curved Spacetime}
\label{sec:quantization}

      There are four basic ingredients in the construction of a quantum field
theory. These are
\begin{itemize}
\item The Lagrangian, or equivalently, the equation of motion of the
classical theory.
\item A quantization procedure, such as canonical quantization or the
path integral approach.
\item The characterization of the quantum states.
\item The physical interpretation of the states and of the observables.
\end{itemize}

In flat spacetime, Lorentz invariance plays an important role in each of 
these steps. For example, it is a guide which generally allows us to identify 
a unique vacuum state for the theory. However, in curved spacetime, we
do not have Lorentz symmetry. This is not a crucial problem in the first two
steps listed above. The formulation of a classical field theory and its
formal quantization may be carried through in an arbitrary spacetime. The
real differences between flat space and curved space arise in the latter
two steps. In general, there does not exist a unique vacuum state in a curved
spacetime. As a result, the concept of particles becomes ambiguous, and
the problem of the physical interpretation becomes much more difficult.

    The best way to discuss these issues in more detail is in the context 
of a particular model theory. Let us consider a real, massive scalar field
for which the Lagrangian density is  
\begin{equation}
{\cal L}=  - \frac{1}{2} (\partial _\alpha \varphi \partial ^\alpha \varphi +
          m^2 \varphi^2 + \xi R \varphi^2).
\end{equation}
(We adopt the  $(+++)$  sign conventions  of Misner, Thorne, and 
Wheeler \cite{MTW}. In particular, the metric signature will be $(- +++)$.
Unless otherwise noted, units in which $G=c=\hbar=1$ are used.) 
The corresponding wave equation is
\begin{equation}
\Box \varphi - m^2\varphi - \xi R\varphi =0. \label{eq:KG}
\end{equation}
Here  $m$ is the mass,   $R$ is the scalar curvature, and $\xi$ is a new coupling constant.
There are two popular choices for $\xi$: minimal coupling ($\xi =0$) and
conformal coupling ($\xi = {1}/{6}$). The former leads to the simplest
equation of motion, whereas the latter leads to a theory which is conformally
invariant in four dimensions in the massless limit. 
For our purposes, we need not settle
this issue, but rather regard $\xi$ on the same footing as $m$, as a parameter
which specifies our theory. Note that here $\Box$ denotes the generally
covariant d'Alembertian operator, $\Box = \nabla_\mu \,\nabla^\mu $.

    A useful concept is that of the {\it inner product} of a pair of solutions
of the generally covariant Klein-Gordon equation, Eq.(\ref{eq:KG}). 
It is defined by
\begin{equation}
(f_1,f_2)=i\int (f^*_2 \, {\mathop{\partial_\mu}\limits^\leftrightarrow } \,f_1) d\Sigma^\mu =
i\int (f^*_2 \, \partial_\mu \,f_1- f_1 \, \partial_\mu \,f^*_2 ) d\Sigma^\mu \,,
\end{equation}
where $d\Sigma^\mu = d\Sigma\, n^\mu$, with $d\Sigma$ being the volume element
in a given spacelike hypersurface, and $n^\mu$ being the timelike unit vector
normal to this hypersurface. The crucial property of the inner product
is that it is independent of the choice of hypersurface. That is, if
$\Sigma_1$ and $\Sigma_2$ are two different, non-intersecting hypersurfaces,
then 
\begin{equation}
(f_1,f_2)_{\Sigma_1} = (f_1,f_2)_{\Sigma_2}. \label{eq:inner}
\end{equation}
The proof of this property is straightforward. We assume that $f_1$ and
$f_2$ are both solutions of Eq. (\ref{eq:KG}). Furthermore, 
if the space is such that
the hypersurfaces are non-compact, we assume that these functions vanish at
spatial infinity. Let $V$ be the four-volume bounded by  $\Sigma_1$ and 
$\Sigma_2$, and, if necessary, time-like boundaries on which $f_1 = f_2=0$.
Then we may write
\begin{equation}
(f_1,f_2)_{\Sigma_2} - (f_1,f_2)_{\Sigma_1} =
i\oint_{\partial V} (f^*_2 \,
{\mathop{\partial_\mu}\limits^\leftrightarrow } \,f_1)
d\Sigma^\mu =
i\int_{V}\nabla_\mu(f^*_2 \,
{\mathop{\partial_\mu}\limits^\leftrightarrow } \,f_1) dV,
\end{equation}
where the last step follows from the four dimensional version of Gauss' law,
and $dV$ is the four dimensional volume element.
However, we may write this integrand as
\begin{eqnarray}
\nabla_\mu(f^*_2 \,{\mathop{\partial_\mu}\limits^\leftrightarrow } \,f_1) 
&=&\nabla_\mu(f^*_2 \partial_\mu \,f_1  -f_1 \partial_\mu \, f^*_2) =
f^*_2 \Box f_1 - f_1 \Box f^*_2 \nonumber \\
&=&-f^*_2 (m^2 +\xi R) f_1 + f_1 (m^2 +\xi R) f^*_2 =0.
\end{eqnarray}
Thus Eq. (\ref{eq:inner}) is proven. 

     The quantization of a scalar field in a curved spacetime may be 
carried out by canonical methods. Choose a foliation of the spacetime
into spacelike hypersurfaces. Let $\Sigma$ be a particular hypersurface
with unit normal vector $n^\mu$ labelled by a constant value of the time
coordinate $t$. The derivative of $\varphi$ in the normal direction is
$\dot \varphi = n^\mu\, \partial_\mu \varphi$, and the canonical momentum
is defined by
\begin{equation}
\pi = {\frac{\delta \cal L}{\delta \dot \varphi}}\quad .
\end{equation}
We impose the canonical commutation relation
\begin{equation}
[\varphi({\bf x},t), \pi({\bf x}',t)] = i\delta({\bf x},{\bf x}'),
\label{eq:CCR}
\end{equation}
where $\delta({\bf x},{\bf x'})$ is a delta function in the hypersurface
with the property that 
\begin{equation}
\int \delta({\bf x},{\bf x}') d\Sigma =1 \, .
\end{equation}

     Let $\{ f_j \}$ be a complete set of positive norm [$(f_j,f_j) >0$]
 solutions of Eq. (\ref{eq:KG}). 
Then $\{ f^*_j \}$ will be a complete set of negative norm solutions,
and $\{ f_j, f^*_j \}$ form a complete set of solutions of the wave equation
in terms of which we may expand an arbitrary solution. Write the field
operator $\varphi$ as a sum of annihilation and creation operators:
\begin{equation}
\varphi = \sum_j (a_j f_j + a^\dagger_j f^*_j),
\end{equation}
where $[a_j, a^\dagger_{j'}] = \delta_{j,j'}$ follows from Eq.~\eqref{eq:CCR}. This expansion defines a
vacuum state $|0\rangle$ such that $a_j |0\rangle=0$. In flat spacetime,
we take our positive norm solutions to be positive frequency solutions,
$f_j \propto e^{-i\omega t}$. Regardless of the Lorentz frame in which
$t$ is the time coordinate, this procedure defines the same, unique
Minkowski vacuum state.

     In curved spacetime, the situation is quite different. There is, 
in general,
no unique choice of the $\{ f_j \}$, and hence no unique notion of the
vacuum state. This means that we cannot identify what constitutes a state
without particle content, and the notion of ``particle'' becomes
ambiguous. One possible resolution of this difficulty is to choose some
quantities other than particle content to label quantum states. Possible
choices might include local expectation values \cite{algebraic}, 
such as $\langle \varphi
\rangle$, $\langle \varphi^2 \rangle$, etc.  In the particular case of
an asymptotically flat spacetime, we might use the particle content in
an asymptotic region. Even this characterization is not unique. However,
this non-uniqueness is an essential feature of the theory with physical
consequences, namely the phenomenon of particle creation, which we will
now discuss.

\section{Particle Creation by Gravitational Fields}
\label{sec:particle-creation}

\subsection{Bogolubov Transformations}
\label{sec:Bogolubov}

    Let us consider a spacetime which is asymptotically flat in the past
and in the future, but which is non-flat in the intermediate region.
Let $\{ f_j \}$ be positive frequency solutions in the past (the 
``in-region''), and let $\{ F_j \}$ be positive frequency solutions in 
the future (the ``out-region'').  We may choose these sets of solutions
to be orthonormal, so that
\begin{eqnarray}
&(f_j,f_{j'})= (F_{j},F_{j'})=\delta _{jj'}& \nonumber \\
&(f_j^*,f_{j'}^*)= (F_{j}^*,F_{j'}^*)= -\delta _{jj'}& \nonumber \\
&(f_j,f_{j'}^*)= (F_{j},F_{j'}^*)= 0.&  \label{eq:ortho}
\end{eqnarray}
Although these functions are defined by their asymptotic properties in
different regions, they are solutions of the wave equation everywhere
in the spacetime. We may expand the in-modes in terms of the out-modes:
\begin{equation}
f_j=\sum\limits_k {(\alpha _{jk}}F_k + \beta _{jk}F_k^*).
\label{eq:in-out}
\end{equation}
Inserting this expansion into the orthogonality relations, 
Eq. (\ref{eq:ortho}), leads to the conditions
\begin{equation}
\sum\limits_k {(\alpha _{jk}\alpha _{j'k}^*-\beta _{jk}\beta_{j'k}^*)
= \delta _{jj'}}, \label{eq:alphabeta}
\end{equation}
and
\begin{equation}
 \sum\limits_k (\alpha _{jk}\beta _{j'k}-\beta _{jk}\alpha _{j'k})=0.
\end{equation}
The inverse expansion is 
\begin{equation}
F_k=\sum\limits_j {(\alpha _{jk}^*}f_j - \beta _{jk}f_j^*).
\end{equation}

The field operator, $\varphi$, may be expanded in terms of either the
$\{ f_j \}$ or the $\{ F_j \}$:
\begin{equation}
  \varphi = \sum\limits_j (a_j f_j + a_j^\dagger f_j^*)
  =\sum\limits_j (b_j F_j + b_j^\dagger F_j^*).
\end{equation}
The $a_j$ and $a_j^\dagger$ are annihilation and creation operators,
respectively, in the in-region, whereas the $b_j$ and $b_j^\dagger$
are the corresponding operators for the out-region. The in-vacuum state
is defined by $a_j|0\rangle_{in}=0, \; \forall j,$ and describes the
situation when no particles are present initially. The out-vacuum state
is defined by $b_j|0\rangle_{out}=0, \; \forall j,$ and describes the
situation when no particles are present at late times. Noting that
$a_j = (\varphi,f_j)$ and $b_j = (\varphi,F_j)$, we may expand the
two sets of creation and annihilation operator in terms of one another
as 
\begin{equation}
a_j=\sum\limits_k (\alpha _{jk}^*b_k-\beta _{jk}^* b_k^\dagger),
                                      \label{eq:Bogo1}
\end{equation}
or 
\begin{equation}
b_k=\sum\limits_j (\alpha _{jk} a_j + \beta _{jk}^* a _j^\dagger).
                                       \label{eq:Bogo2}
\end{equation}
This is a Bogolubov transformation, and the $\alpha_{jk}$ and
$\beta_{jk}$ are called the Bogolubov coefficients.

   Now we are ready to describe the physical phenomenon of particle
creation by a time-dependent gravitational field. Let us assume that
no particles were present before the gravitational field is turned on.
If the Heisenberg picture is adopted to describe the quantum dynamics,
then $|0\rangle_{in}$ is the state of the system for all time. However,
the physical number operator which counts particles in the out-region
is $N_k = b_k^\dagger b_k$. Thus the mean number of particles created
into mode $k$ is 
\begin{equation}
\langle N_k \rangle = {}_{in}\langle 0|b_k^\dagger b_k |0\rangle_{in}
       = \sum\limits_j |{\beta _{jk}}|^2.
 \label{eq:N}      
\end{equation}
If any of the $\beta_{jk}$ coefficients are non-zero, i.e. if
any mixing of positive and negative frequency solutions occurs, then
particles are created by the gravitational field.  The above discussion describes the 
spontaneous creation of bosons, which will be the main interest in this review,
but an analogous procedure may be used to describe fermion creation.

      The most straightforward application of the concepts developed above
is to particle creation by an expanding universe. This phenomenon was
first hinted at in the work of Schr\"odinger \cite{Schrodinger}, 
but was first carefully investigated by Parker \cite{Parker69,Parker71}. 
Let us restrict our attention to the case of a
spatially flat Robertson-Walker universe, for which the metric may be
written as
\begin{equation}
ds^2 = -dt^2 + a^2(t) d{\bf x}^2 = a^2(\eta)\,
       \bigl( - d\eta^2  + d{\bf x}^2\bigr),
       \label{eq:metric}
\end{equation}
where $a$ is the scale factor. We may use either the comoving time $t$
or the conformal time $\eta$, but the solutions of the wave equation are
simpler in terms of the latter. The positive norm solutions of 
Eq. (\ref{eq:KG}) in this metric may be taken to be 
\begin{equation}
f_{\bf k}({\bf x},\eta)={\frac{e^{i{\bf k\cdot x}}}{a(\eta)\sqrt {(2\pi)^3}}}\,\, \chi_k(\eta), 
\label{eq:f-def}
\end{equation}
where $\chi_k(\eta)$ satisfies 
\begin{equation}
{\frac{d^2\chi_k}{d\eta^2}} + [ {k^2 + V(\eta)} ]\chi_{k} =0 \, ,
                                                 \label{eq:chieq}
\end{equation}
with
\begin{equation}
V(\eta) \equiv a^2(\eta) 
        \biggl[m^2 + \Bigl( \xi - \frac{1}{6}\Bigr) R(\eta)\biggr].
        \label{eq:V}
\end{equation}
The norm of $f_{\bf k}$ being equal to one is equivalent to the Wronskian 
condition
\begin{equation}
\chi_k {\frac{d \chi_k^*}{d\eta}} - \chi_k^* {\frac{d \chi_k}{d\eta}} =i.
\end{equation}

\begin{figure}[htbp]
\includegraphics[scale=0.4]{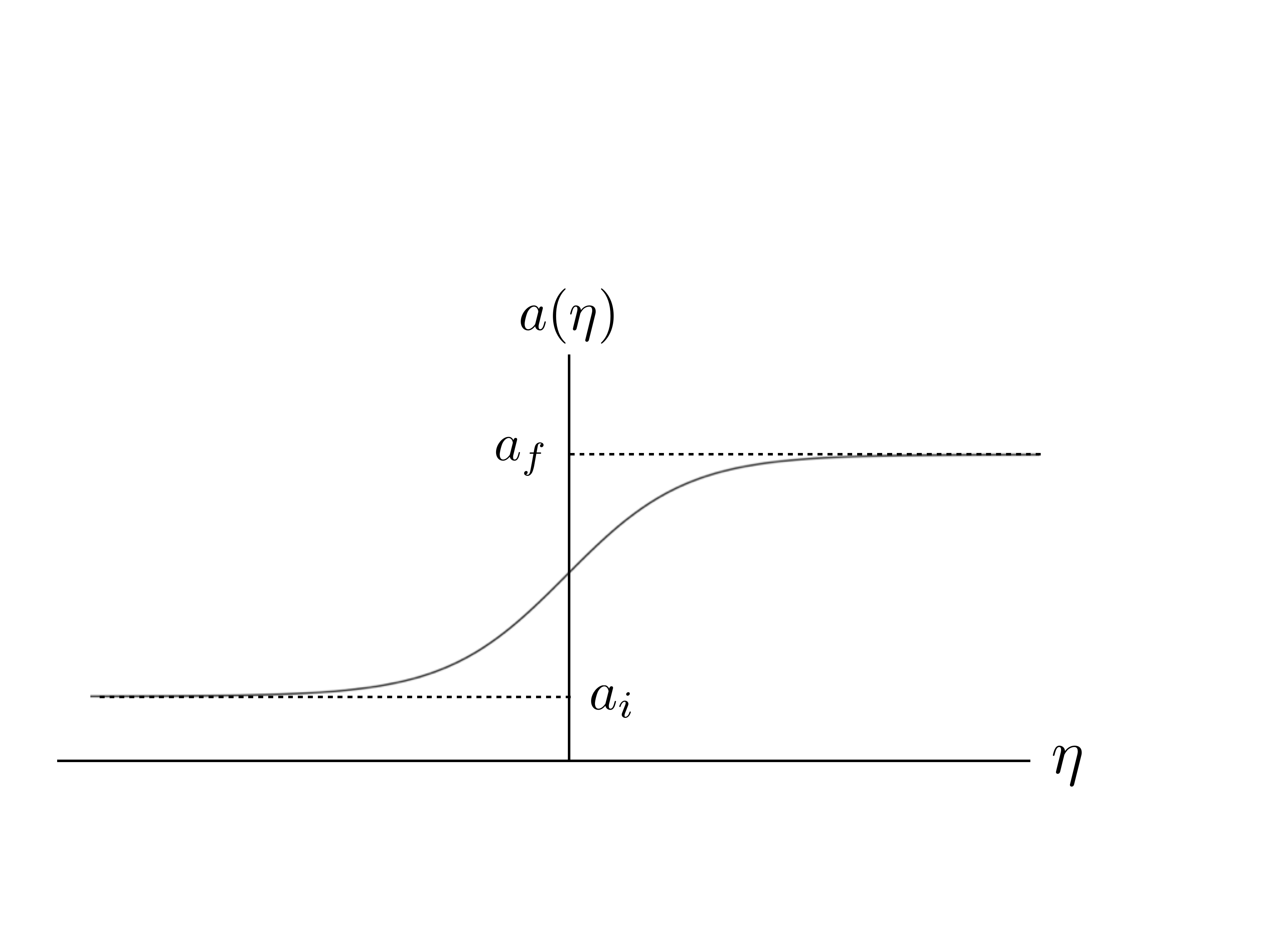}
\caption{Here the scale factor $a(\eta)$ for an asymptotically bounded expansion of the universe is illustrated,
In the past, the in-region, spacetime is flat and $a=a_i$. The universe then expands for a finite interval of conformal
time, $\eta$. In the future, the out-region, spacetime is again flat and  $a=a_f$. Often we set $a_f = 1$ for convenience.}
\label{fig:boundedexpansion}
\end{figure}

Let us consider the idealized situation in which the universe is static 
both in the past and in the future, as illustrated in Fig.~\ref{fig:boundedexpansion}. 
In this case, we have the necessary
asymptotically flat regions needed to define in and out vacua. Let
 \begin{equation}
a(\eta)  \rightarrow  \begin{cases} 
a_i & \eta \rightarrow -\infty \\  a_f &   \eta \rightarrow \infty 
\end{cases}\,
\label{eq:a-static}
\end{equation} 
where $a_i$ and $a_f$ are constants. Let $ \chi_k^{(in)}(\eta)$ be a solution of
Eq.~\eqref{eq:chieq} which is pure positive frequency in the past:
 \begin{equation}
 \chi_k^{(in)}(\eta) \sim {\frac{e^{-i\omega_{in} \eta}}{\sqrt{2\omega_{in}}}},
 \qquad \eta \rightarrow -\infty\,.
 \label{eq:chipast}
 \end{equation}
Similarly,  $ \chi_k^{(out)}(\eta)$ is pure positive frequency in the future:
\begin{equation}
 \chi_k^{(out)}(\eta) \sim {\frac{e^{-i\omega_{out} \eta}}{\sqrt{2\omega_{out}}}},
 \qquad \eta \rightarrow \infty\,.
   \label{eq:chifuture}
\end{equation}
Here 
\begin{equation}
 \omega_{in} = \sqrt{k^2 + a_i^2\, m^2}\,,
 \label{eq:omega-in}
 \end{equation}
 and
\begin{equation}
 \omega_{out} = \sqrt{k^2 + a_f^2\, m^2}\,.
  \label{eq:omega-out}
 \end{equation}
When we include the spatial dependence, the positive frequency modes in the past are given by
Eq.~\eqref{eq:f-def} with $\chi_k(\eta) = \chi_k^{in}(\eta)$, and those which are positive frequency
in the future, $F_{\bf k}({\bf x},\eta)$, are given by the same relation, but with $\chi_k(\eta) = \chi_k^{out}(\eta)$.

Note that although these modes are defined by their asymptotic forms in the past or future, both sets are
valid solutions for all $\eta$, and hence one set may be expanded in terms of the other. Let
 \begin{equation}
  \chi_k^{(in)}(\eta) = \frac{a_i}{a_f}\, [\alpha_k\, \chi_k^{(out)}(\eta) + \beta_k\, \chi_k^{(out)}(\eta)^*]\,
  \label{eq:chi-alphabeta}
 \end{equation}
for some constants $\alpha_k$ and $\beta_k$. This leads to the relation
 \begin{equation}
 f_{\bf k}({\bf x},\eta)= \alpha_k\,  F_{\bf k}({\bf x},\eta) + \beta_k \, F^*_{\bf k}({\bf x},\eta) \,.
 \end{equation}
This is just Eq.~\eqref{eq:in-out}, where  the Bogolubov coefficients are given 
by $\alpha_{\bf k k'} = \alpha_k \delta_{\bf k k'}$ and $\beta_{\bf k k'} = \beta_k \delta_{\bf k, -k'}$.
The condition Eq.~\eqref{eq:alphabeta} becomes
 \begin{equation}
 |\alpha_k|^2 -  |\beta_k|^2 = 1\,,
 \label{eq:alphabeta2}
 \end{equation}
 and the mean number of particles created in a given mode, from Eq.~\eqref{eq:N}, is now $\langle N_k \rangle =  |\beta_k|^2$.
 The simple forms of the Bogolubov coefficients in a spatially flat universe follow from the spatial
 translation symmetry of the metric, Eq.~\eqref{eq:metric}. This means that three-momentum is conserved,
 even though energy is no longer conserved on a time-dependent background.  Indeed, it is the lack of energy conservarion 
 which allows particle creation from the vacuum in an expanding universe. It also allows other processes which are forbidden
 in flat spacetime, including new decay channels for unstable particles~\cite{Ford82,Lankinen18}. 
 
 If we quantize the field $\varphi$ using periodic boundary condition in a box with coordinate volume ${\cal V}$,
 then the wave number ${\bf k}$ takes discrete values, and in the limit of large  ${\cal V}$, 
 \begin{equation}
 \sum_{\bf k} \rightarrow \frac{\cal V}{(2 \pi)^3} \int d^3 k\,,
 \label{eq:large-V}
 \end{equation}
just as in flat spacetime. However, the physical volume of the box in the out-region is $a_f^3\, {\cal V}$, so
the total number of created particle per unit volume becomes
\begin{equation}
n = \frac{1}{(2 \pi \, a_f)^3}\,  \int d^3 k\,  |\beta_k|^2\,.
\label{eq:n} 
 \end{equation}
 The density of created particle per unit volume and per unit wave number interval is
 \begin{equation}
\frac{dn}{dk} = \frac{1}{2 \pi^2 \, a_f^3}\,   k^2\,  |\beta_k|^2\,.
\label{eq:dn} 
 \end{equation}
The energy of a particle in the final region is
 \begin{equation}
 E_{out} = \frac{\omega_{out}}{a_f} = \sqrt{ \left(\frac{k}{a_f}\right)^2 +m^2} \,.
 \end{equation}
 Note that in $\hbar =1$ units, $k$ has the units of momentum, and $k/a_f$ is the physical momentum of
 the particle as measured by an observer at rest in the cosmological frame  in the final region.
 The energy density of created particles is
 \begin{equation}
 u = \frac{1}{(2 \pi \, a_f)^3}\,  \int d^3 k\,E_{out} \,  |\beta_k|^2\,.
 \label{eq:u} 
 \end{equation}
In the massless limit, $m=0$, this becomes
\begin{equation}
 u_0 = \frac{1}{(2 \pi)^3 \, a_f^4}\,  \int d^3 k\, k \,  |\beta_k|^2\,.
 \label{eq:u0} 
 \end{equation}
The extra factor of $a_f$ in the denominator of Eq.~\eqref{eq:u0}, as compared to Eq.~\eqref{eq:n},
arises from the redshifting of the energy of massless particles by the cosmological expansion.

Note that conformally invariant massless particles are not created by cosmological
expansion in a conformally flat spacetime, such as that of Eq.~\eqref{eq:metric}, as was
first noted by Parker~\cite{Parker69}. We can see this in the case of the scalar field; if we set
$m=0$ and $\xi = 1/6$ in Eq.~\eqref{eq:V}, then $V =0$ and the solutions of Eq.~\eqref{eq:chieq}
become proportional to ${\rm e}^{\pm i k \eta}$. This means that a positive frequency solution in the past
remains a positive frequency solution, so that $\beta_k =0$ for all modes and no particles are created.
Thus particle creation requires a nonzero mass or non-conformal coupling to the curvature,
$\xi \not= 1/6$, or both. For most choices of the scale factor, $a(\eta)$, it can be difficult to find
exact solutions to Eq.~\eqref{eq:chieq}. However, there are a few exactly soluble models which we 
will now consider.

\subsection{Specific Models}
\label{sec:specific}

\subsubsection{The Bernard and Duncan Model: Effects of nonzero mass}
\label{sec:BD}
 One exact solution which illustrates the effects of nonzero mass was given by Bernard and 
 Duncan~\cite{BeDu}, and is also reviewed in Sec.~3.4 of  Ref.~\cite{BD}. This model was 
 originally formulated in a two-dimensional spacetime, but can be easily extended to a four-dimensional model, 
 which will be done here. Consider a universe with the metric  Eq.~\eqref{eq:metric}, where
  \begin{equation}
 a^2(\eta) = \frac{1}{2}\, [1 + a_i^2 +(1-a_i^2) \tanh(\rho \eta)]\,
 \label{eq:BD-scale}
 \end{equation}
This describes a universe which expands from an initial scale factor of $a_i < 1$ to a final
scale factor of $a_f =1$, as illustrated in Fig.~\ref{fig:boundedexpansion}  . For the case of conformal coupling, $\xi =1/6$,
Eq,~\eqref{eq:chieq} becomes
 \begin{equation}
 {\frac{d^2\chi_k}{d\eta^2}} + \left\{ k^2 +  \frac{1}{2} m^2\, \left[1 + a_i^2 +(1-a_i^2) \tanh(\rho \eta)\right]\right\}\chi_{k} =0 \, .
                                                 \label{eq:chieq-BD}
 \end{equation} 

This equation may be solved in terms of hypergeometric functions. The solution which is pure positive 
frequency in the past, and has the asymptotic form given in Eq.~\eqref{eq:chipast}, is
 \begin{equation}
 \chi^{in}_k(\eta) =    \frac{1}{\sqrt{2\omega_{in}}}\, \exp\left\{-i \omega_{+}   \eta - i \frac{\omega_{-}}{ \rho} \ln[2 \cosh(\rho \eta) \right\}\;
 F\left(1+\frac{i \omega_{-}}{\rho}, \frac{i \omega_{-}}{\rho}; 1-\frac{i \omega_{in}}{\rho}; 
\frac{1}{2} [1+ \tanh(\rho \eta)] \right)\,,
\label{eq:chi-in}
 \end{equation}
 where $\omega_\pm = (\omega_{out} \pm \omega_{in})/2$. Here $F$, sometimes denoted by ${}_2F_1$, is a hypergeometric funcion.
The corresponding solution which  is pure positive  frequency in the future and has the asymptotic form in Eq.~\eqref{eq:chifuture} is
 \begin{equation}
 \chi^{out}_k(\eta) =  \frac{1}{\sqrt{2\omega_{out}}}\, \exp\left\{-i \omega_{+}   \eta - i \frac{\omega_{-}}{ \rho} \ln[2 \cosh(\rho \eta) \right\}\;
 F\left(1+\frac{i \omega_{-}}{\rho}, \frac{i \omega_{-}}{\rho}; 1 + \frac{i \omega_{out}}{\rho}; 
\frac{1}{2} [1- \tanh(\rho \eta)] \right)\,.
\label{eq:chi-out}
 \end{equation}
 
 The two solutions $ \chi^{in}_k(\eta)$ and $ \chi^{out}_k(\eta)$ are related by the linear transformation
 properties of the hypergeometric function, which leads to a relation of the form of Eq.~\eqref{eq:chi-alphabeta}.
Here
 \begin{equation}
 \alpha_k =  \frac{1}{a_i}\, \sqrt{\frac{\omega_{out} }{\omega_{in} }}\, \frac{\Gamma(1-i \omega_{in}/\rho) \, \Gamma(-i \omega_{out}/\rho)}
 {\Gamma(1- i \omega_{+}/\rho) \, \Gamma( - i \omega_{+}/\rho)} \,,
 \label{eq:alpha-BD}
 \end{equation}
 and 
  \begin{equation}
 \beta_k = \frac{1}{a_i}\, \sqrt{\frac{\omega_{out} }{\omega_{in} }}\, \frac{\Gamma(1-i \omega_{in}/\rho) \, \Gamma(i \omega_{out}/\rho)}{\Gamma(1+i \omega_{-}/\rho) \, \Gamma(i \omega_{-}/\rho)} \,.
 \label{eq:beta-BD}
 \end{equation}
 This leads to
  \begin{equation}
  | \beta_k|^2 = \frac{  \sinh^2(\pi \omega_{-}/\rho)}{ a_i^2\, \sinh(\pi\,  \omega_{out}/\rho)\, \sinh(\pi \omega_{in}/\rho)}\,.
 \label{eq:N-BD}
 \end{equation}
 These particles each have energy $\omega_{out} = \sqrt{k^2 + m^2}$ in the out-region. 
 
 In the limit of small mass, so $ m \ll k a_i < k$, we have
 \begin{equation}
\omega_{out} \approx k + \frac{m^2}{2 k^2} +\cdots\,, \quad \omega_{in} \approx k + \frac{a_i^2 \,m^2}{2 k^2} +\cdots\,, \quad 
\omega_{-} \approx \frac{m^2}{ 4 \,k}(1-a_i^2)\,,
 \end{equation}
so
\begin{equation}
  | \beta_k|^2 \approx \frac{ \pi^2 (1-a_i^2) \, m^4}{ 16\, k^2\, a_i^2 \,\rho^2\, \sinh^2(\pi \, k/\rho)} \,.
  \end{equation}
 Thus the number of created particles vanishes as $m^4$ when $m \rightarrow 0$. If we also assume that $k \agt \rho$, then
 \begin{equation}
  | \beta_k|^2 \approx \frac{ \pi^2 (1-a_i^2)^2 \, m^4}{ 4 k^2\, a_i^2 \,\rho^2}  \: {\rm e}^{-2 \pi k/\rho}   \,.
   \label{eq:N-BD2}
  \end{equation} 
In the limit of large mass, so $m \gg k > a_i k$, Eq.~\eqref{eq:N-BD} becomes  
 \begin{equation}
  | \beta_k|^2 \approx \frac{1}{a_i^2} \;  {\rm e}^{-2\pi m a_i/\rho}\,.
   \label{eq:N-BD3}
 \end{equation}
 This limit describes particles which are created at rest in the cosmological frame, but with an exponentially suppressed probability.
 
 A somewhat different limit may be explored by assuming that $\rho$ becomes small with $k$, $m$, and $a_i$ fixed, so that
 $\omega_{in}$, $\omega_{out}$, and $\omega_{-}$, are all large compared to  $\rho$. This is the limit of slow expansion, where we find  
 \begin{equation}
  | \beta_k|^2 \approx \frac{1}{a_i^2} \;  {\rm e}^{-2\pi \, \omega_{in}/\rho}\,.
   \label{eq:N-BD4}
 \end{equation}  
 If we use Eq.~\eqref{eq:omega-in}, and assume that $k \alt  m a_i$, then we again obtain Eq.~\eqref{eq:N-BD3}. However,
 if we now assume that $k \agt  m a_i$, then we find
 \begin{equation}
  | \beta_k|^2 \approx \frac{1}{a_i^2} \;  {\rm e}^{-2\pi \, k/\rho}\,.
   \label{eq:N-BD4a}
 \end{equation}  
 This result contains the same exponential factor as does Eq.~\eqref{eq:N-BD2}, although with a different prefactor, arising from 
 assuming somewhat different limits. Both forms show that the creation of very energetic particles is exponentially suppressed.

  Note that in the two-dimensional version of this model described in Refs.~\cite{BeDu,BD},  the factors of $a_i$ in the final
  results, such as Eq.~\eqref{eq:N-BD}, do not appear. This is because in two spacetime dimensions, the factor of $1/a(\eta)$
  which appears in four dimensional solutions of the Klein-Gordon equation, e.g. Eq.~\eqref{eq:f-def}, does not appear.  
  In four spacetime dimensions, the fact that $ | \beta_k|^2 \propto   a_i^{-2}$ leads to enhanced particle creation when 
  $a_i \ll 1$ and the fractional change in scale factor is large.  
  
  The total number density and energy density of the created particles are given by  Eqs.~\eqref{eq:n}  and \eqref{eq:u},
  respectively, with $a_f = 1$. For the purposes of an estimate, let us use Eq.~\eqref{eq:N-BD2}. Although this form does not
  hold over the entire range of integration, it should be a fair approximation for $k \approx \rho$, which is expected to give the
  dominant contribution to the integrals. In addition, we use $E_{out}  \approx k$. This leads  to
   \begin{equation}
 n \approx  \frac{  (1-a_i^2)^2 \, m^4}{ 16 \pi \, a_i^2 \,\rho}\,,
 \label{eq:n-BD}
 \end{equation}
 and 
  \begin{equation}
 u \approx  \frac{  (1-a_i^2)^2 \, m^4}{ 32 \pi^2 \, a_i^2}\,.
 \label{eq:u-BD}
 \end{equation}
 If we take the ratio of these two expressions, we may estimate the average energy of the created particles to be
 \begin{equation}
E_{ave} = \frac{u}{n} \approx 2 \pi\, \rho\,,
\label{eq:E-BD}
 \end{equation}
 so this average particle energy is proportional to the inverse expansion time, $\rho$.

 \subsubsection{The Parker and Toms Model: Effects of non-conformal coupling}
\label{sec:PT}
 
 There is a similar model, given by Parker and Toms~\cite{Parker76,PT}, with statically bounded expansion which provides an exact solution 
 for the creation of massless, minimally coupled scalar particles, and hence of gravitons as will be discussed in Sec.~\ref{sec:gravitons}. 
 If one defines a new time coordinate $\tau$ in the metric Eq.~\eqref{eq:metric}    by $dt = a^3 \, d\tau$, then the metric becomes 
 \begin{equation}
ds^2 = - a^6(\tau) d\tau^2 + a^2(\tau) d{\bf x}^2 \,,
       \label{eq:tau-metric}
\end{equation}
 and the solutions of the massless, minimally coupled Klein-Gordon equation $\Box \varphi =0$ can be written as
 \begin{equation}
f_{\bf k}({\bf x},\tau)= \frac{e^{i{\bf k\cdot x}}}{\sqrt{(2\pi)^3}}\,\, \chi_k(\tau), 
\label{eq:f-tau-def}
\end{equation}
Note that there is no explicit scale factor in this expression, in contrast to Eq.~\eqref{eq:f-def}.
 Now $\chi_k(\tau)$ satisfies the equation
\begin{equation}
\frac{d^2\chi_k}{d\tau^2} +  k^2 \, a^4(\tau) \,\chi_{k} =0 \, .
                                                 \label{eq:chi-tau}
\end{equation}

 The scale factor used by Parker and Toms can be expressed in a form analogous to Eq.~\eqref{eq:BD-scale}
  \begin{equation}
 a^4(\tau) = \frac{1}{2}\, [1 + a_i^4 +(1-a_i^4) \tanh(\rho \tau)]\,,
 \label{eq:PT-scale}
 \end{equation}
and also expands from an initial value of $a_i < 1$ to a final value of $a_f =1$, as illustrated in Fig.~\ref{fig:boundedexpansion},
but now as a function of $\tau$. Equation~\eqref{eq:chi-tau} becomes
\begin{equation}
 {\frac{d^2\chi_k}{d\tau^2}} + \frac{1}{2}  k^2 \,  \left[1 + a_i^4 +(1-a_i^4) \tanh(\rho \tau)\right]\, \chi_{k}(\tau) =0 \, ,
                                                 \label{eq:chieq-PT}
 \end{equation} 
which is the same as Eq.~\eqref{eq:chieq-BD} if we make the substitutions
 \begin{equation}
 k \rightarrow 0\,, \quad m\rightarrow k\, \quad a_i^2   \rightarrow a_i^4\,, \quad  \eta \rightarrow \tau
 \end{equation}
 in the latter.
 As a consequence, the solutions for the in and out modes, and also the Bogolubov coefficents, may be found 
 from those in Sect.~\ref{sec:BD} by this substitution, except here there will be no overall factors of $a_i$.
 Now the Bogolubov coefficents are given by Eqs.~\eqref{eq:alpha-BD} and \eqref{eq:beta-BD} without the
 leading $1/a_i$ factor, and where now
 \begin{equation}
 \omega_{in} = k\, a_i^2\,,  \quad \omega_{out} = k\,,
 \label{eq:subs}
 \end{equation}
 so here $\omega_\pm = k (1 \pm a_i^2)/2$.
 Now we have the result
  \begin{equation}
  | \beta_k|^2 = \frac{  \sinh^2[\pi k (1-a_i^2)/(2\rho)]}{ \sinh(\pi\, k/\rho)\, \sinh(\pi k\, a_i^2/\rho)}\,.
 \label{eq:N-PT}
 \end{equation}
 In the high frequency limit, $k > k\, a_i^2 \agt \rho$, we have
  \begin{equation}
  | \beta_k|^2 \approx {\rm e}^{-2 \pi\, k \,a_i^2/\rho}\,.
 \label{eq:N-PT2}
 \end{equation}
 Equations~\eqref{eq:N-BD2}, \eqref{eq:N-BD3}, \eqref{eq:N-BD4}, and \eqref{eq:N-PT2} all illustrate that the creation rate of
 particles with energies larger than the inverse expansion time, $\rho$, is exponentially suppressed. Parker~\cite{Parker76} has
 noted that these exponential factor  are similar to those which appear in thermal spectrum at finite temperature. Note, however,
 that here the particles are created in a pure quantum state and hence have zero entropy until decoherence occurs. This issue  
 will be treated in Sect.~\ref{sec:state-entropy}.  Several other authors have discussed cases where an approximately thermal
 spectrum of particles is created~\cite{HuRaval96,Koks97,Mersini98,Biswas02}.
 
 As in the previous subsection, we may use Eq.~\eqref{eq:N-PT2} to obtain estimates for the total number
 and energy density of the created particles. The results are
  \begin{equation}
 n \approx \frac{\rho^3}{8 \pi^5 \, a_i^6}\,
 \label{eq:n-PT}
 \end{equation}
 and
  \begin{equation}
 u \approx \frac{3 \rho^4}{16 \pi^6 \, a_i^8} \,.
 \label{eq:u-PT}
 \end{equation}
 The associated estimate for the mean energy of the created particles is
  \begin{equation}
E_{ave} = \frac{u}{n} \approx \frac{3 \rho}{2 \pi \, a_i^2}  \,.
\label{eq:E-PT}
 \end{equation}
 The inverse powers of $a_i$ in the above expressions indicate that a large change in scale factor, where 
 $a_i \ll 1$, is especially effective in creating massless minimally coupled scalars and gravitons. 
 
 \subsection{A Perturbative Approach}
\label{sec:Pert}

   It is difficult to solve Eq.~(\ref{eq:chieq}) for the mode 
functions in all but a few examples, such as those just discussed. However, there is a perturbative
method, developed by Zeldovich and Starobinsky \cite{ZS} and by Birrell and
Davies~\cite{BD2}, which is often useful. The first step is to rewrite Eq.~\eqref{eq:chieq}
as
\begin{equation}
{\frac{d^2\chi_k}{d\eta^2}} + [ \omega_{in}^2 + U(\eta) ]\chi_{k} =0 \, ,
                                                 \label{eq:chieq-2}
\end{equation}
where
\begin{equation}
U(\eta) =V(\eta)-a_i^2\, m^2 = m^2[a^2(\eta) -a_i^2] 
        +a^2(\eta)\, \Bigl( \xi - \frac{1}{6}\Bigr) R(\eta)\,.
        \label{eq:U}
\end{equation}
Note that $U(\eta) \rightarrow 0$ as $\eta \rightarrow -\infty$.
Next we rewrite Eq.~\eqref{eq:chieq-2} as an integral equation:
\begin{equation}
\chi_k(\eta) = \chi_k^{(in)}(\eta) - \omega_{in}^{-1} \int_{-\infty}^\eta
             U(\eta')\;\sin \, [\omega_{in}(\eta-\eta')]\;\chi_k(\eta') d\eta' \,,
\end{equation}
where here  $\chi_k^{(in)}(\eta)$ is understood to be the asymptotic form given in Eq.~\eqref{eq:chipast}.
This integral equation is equivalent to Eq.~(\ref{eq:chieq-2}) 
plus the boundary condition that $\chi_k(\eta) \sim \chi_k^{(in)}(\eta)$
for $\eta \rightarrow -\infty$.
We now wish to assume that $U$ is sufficiently small that we
may iterate this integral equation to lowest order by replacing
$\chi_k(\eta')$ by $\chi_k^{(in)}(\eta')$ in the integrand. If we
compare the resulting formula in the limit $\eta \rightarrow \infty$  with 
the right-hand side of Eq.~\eqref{eq:chi-alphabeta},  
the Bogolubov coefficients may be read off:
\begin{equation}
\alpha_k \approx   \frac{a_f}{a_i}\,  \left[1 - \frac{i}{2\omega_{in}} \, \int_{-\infty}^\infty
             U(\eta)\, d\eta \right]\,,
\end{equation}
and
\begin{equation}
\beta_k \approx    \frac{i\, a_f}{2  a_i\, \omega_{in}} \,  \int_{-\infty}^\infty
            e^{-2i\omega_{in} \eta}\; U(\eta)\, d\eta.     \label{eq:beta-pert}
\end{equation}
 
 We can see that the above procedure is a first order expansion in $m^2$ and in
 $( \xi - \frac{1}{6})\, R$, or more precisely, the dimensionless ratios of these two quantities to
 the inverse squared expansion time scale. In the limit that $m^2 = ( \xi - \frac{1}{6})\, R =0$, there is no particle
 creation, as expected. Otherwise, Eq.~\eqref{eq:beta-pert} tells us that the particle creation will be peaked
 at the values of $\omega_{in}$ associated with the peak of the Fourier transform of $U(\eta)$, which is of the 
 order of the characteristic inverse  time scale, in conformal time, of the expansion. Strictly, the perturbative
 approach assumes that $U$ is always small, which requires that the scale factor not change by large factors
 if $m \not=0$, in which case $\omega_{out} \approx \omega_{in}$. 
 
 We can compare the perturbative result with the exact solution in the Bernard and Duncan model.
 In this case, the squared scale factor is given by Eq.~\eqref{eq:BD-scale} and $\xi =1/6$. The integral in
 Eq.~\eqref{eq:beta-pert} may be performed explicitly. The result contains a term proportional to $\delta(\omega_{in})$,
 which would seem to describe created particles with zero energy, and which does not appear in the exact solution. If we
 assume that this term should not be present, and drop it, we then find
  \begin{equation}
    | \beta_k|^2 \approx \frac{ \pi^2 (1-a_i^2)^2 \, m^4}{ 4 \omega_{in}^2\, a_i^2 \,\rho^2}  \: {\rm e}^{-2 \pi \omega_{in}/\rho}   \,,
 \end{equation}
 where we have set $a_f = 1$ and assumed that $\omega_{in} \agt \rho$.
This result agrees with Eq.~\eqref{eq:N-BD2} in the small mass limit, where $\omega_{in} \approx k + O(m^2)$ and
$\omega_{out} \approx \omega_{in}$. 

Let us now consider the case where $m=0$. In this case, the
mean number density becomes
\begin{equation}
n= \frac{(\xi - \frac{1}{6})^2}{16\pi a_f^3} \int_{-\infty}^\infty
                        a^4(\eta)\,R^2(\eta)\,d\eta,
                        \label{eq:Npert}
\end{equation}
and the energy density becomes
\begin{eqnarray}
u = - \frac{(\xi - \frac{1}{6})^2}{32\pi^2 a_f^4} \int_{-\infty}^\infty d\eta_1
    \int_{-\infty}^\infty d\eta_2 &{}& \!\!\!\!\!\!\!\!\! 
    \biggl\{ \ln(|\eta_1-\eta_2|\mu)
    \frac{d}{d\eta_1 }\Bigl[a^2(\eta_1)\,R(\eta_1)\Bigr] \nonumber \\
    &\times & \frac{d}{d\eta_2}\Bigl[a^2(\eta_2)\,R(\eta_2)\Bigr] \biggr\}\,.
     \label{eq:rhopert}
\end{eqnarray}
Here $\mu$ is an arbitrary quantity with the dimensions of mass; $u$
is independent of $\mu$ provided that $a^2(\eta)\,R(\eta) \rightarrow 0$
as $\eta \rightarrow \pm \infty$.
The approximation which  is being used in the $m=0$ case amounts to perturbation around
the conformally invariant theory in powers of $(\xi - \frac{1}{6})$. Note that here $n$ and $u$
are independent of $a_i$, and we have restored a general value of $a_f$.

\subsection{Estimates of Magnitude of the Particle Creation}
\label{sec:rate-estimates}

The rate of particle creation is not precisely defined, but in the massless case one can  view the
integrand of Eq,~\eqref{eq:Npert} as an approximate number density creation rate per
unit conformal time $\eta$. The energy density creation rate is more complicated, because
Eq,~\eqref{eq:rhopert} involves a double time integration, and because the integrand depends
upon the arbitrary constant $\mu$. However, if one estimates a characteristic energy of the created 
particles from the Fourier transform in Eq.~\eqref{eq:beta-pert}, then this may be combined with
the  approximate number density creation rate to estimate the energy density creation rate.
However, in the case that $m \not= 0$, even the particle number is expressed as a double 
time integral~\cite{BD2}, so there no obvious candidate of a number creation rate.

More generally, we may combine the perturbative method with the exact solutions given in
Secs.~\ref{sec:BD} and \ref{sec:PT} to make some estimates of the number and spectrum
of the created particles. First, Eqs.~\eqref{eq:dn}  and \eqref{eq:beta-pert} tell us that the
number density of particles created into a given interval of wavenumber or energy is proportional
to the square of a Fourier component of the function $U(\eta)$ associated with that energy interval.
Let $\eta_C$ be the characteristic timescale associated with the scale factor $a(\eta)$. For example, in
the case of Eq.~\eqref{eq:BD-scale}, we could take $\eta_C = 1/\rho$. 

We have found in the exact solutions in Secs.~\ref{sec:BD} and ~\ref{sec:PT} that the creation of
of particles with energies large compared to $1/\eta_C$ is exponentially suppressed, as seen in 
Eqs.~\eqref{eq:N-BD2}, \eqref{eq:N-BD3}, \eqref{eq:N-BD4}, and \eqref{eq:N-PT2}. The feature
also appears in the perturbative results in Sec.~\ref{sec:Pert} due to exponential decay of Fourier
tranforms at high frequencies for a wide class of functions. More generally, we can expect the
energy spectrum of the created particles to be peaked near energies $E \propto 1/\eta_C$ , as 
illustrated in Eqs.~\eqref {eq:E-BD} and \eqref{eq:E-PT}. 

The particle creation rate depends upon the cosmological expansion rate, which is in turn determined
by the matter equation of state. The effects of a stiff equation of state have been discussed by
Lankinen and Vilia~\cite{Lankinen2017}.   

Note that this review has focussed on the quantum creation of bosons, especially scalar particles.
The creation of fermions in an expanding universe was treated by Parker~\cite{Parker71}, where
anti-commutation relations replace the commutation relation used here in Sec.~\ref{sec:Bogolubov}.
In general, one can expect the creation rate of fermions to be somewhat less than than for bosons
of similar mass and coupling to curvature. This is due to the effects of the exclusion principle for
fermions.

\subsection{Quantum Creation of Gravitons}
\label{sec:gravitons}

Although there is not yet  a universally accepted quantum theory of gravity, there are  limits in which gravity may be quantized
unambiguously. One such limit is the quantization of tensor perturbations of an expanding universe, such as that of Eq.~\eqref{eq:metric}.
This is  treated, for example, in Ref.~\cite{FordParker:1977}. If one imposes the transverse tracefree gauge, then all of the
gauge freedom is removed, leaving only the two physical degrees of freedom associated with the two independent choices
of polarization. The metric perturbations in this gauge satisfy the scalar wave equation, Eq.~\eqref{eq:KG}, with $m = \xi =0$,
that is the equation for a massless, minimally coupled scalar field~\cite{Lifshitz}. Because of the two polarization degrees of
freedom, we can view gravitons in a spatially flat expanding universe as being equivalent to a pair of massless minimally coupled
fields. A somewhat different approach using Hamiltonian methods was given by Berger~\cite{Berger74}.

A key point is that the wave equation for gravitons, unlike that for photons, is not conformally invariant. The means that in
general gravitons will be created by the cosmological expansion, as was  emphasized by Grishchuk~\cite{Grishchuk:1975}.
For example, the creation of gravitons by the Parker-Toms model of Sect.~\ref{sec:PT} may be obtained by 
multiplying the number density or energy density of  created massless scalar particles by two. Similarly, the perturbative method
of Sect.~\ref{sec:Pert} may be used to obtain the number density or energy density of the created gravitons. In this case, set $\xi=0$
in Eqs.~\eqref{eq:Npert} and \eqref{eq:rhopert}, and then multiply the result by two to account for the two graviton polarization states. 
An explicit example will be treated in Sec.~\ref{sec:late-time}.

\subsection{The Issues of Initial Conditions and Local Definitions of Particle Number}
\label{sec:initial}

Here we have used initial and final flat regions to give asymptotic definitions of particle number. In Sect~\ref{sec:Bunch-Davies},
we will discuss a more natural definition of initial particle number in the context of inflationary cosmology. In general, particle number
is not well defined during the expansion. Expressions such as Eq.~\eqref{eq:Npert} give a total number of created particles
as a time integral, and it is tempting to interpret the integrand as a number creation rate. However, this should be done with caution.
In general, one should not expect the instantaneous number of particles during the expansion to be well defined. Nonzero particle
creation occurs when the particle number definitions in the initial and final regions differ. Certainly particle creation is also linked to the
time dependence of the background and the lack of a conserved energy on such a background. 

There are limits
in which a approximate local definition may be given. When the wavelength of the particle is small compared to the local radius of
spacetime curvature, then the particle is locally in an approximately flat region of spacetime, and particle number is meaningful. 
This may be made more precise in the context of an adiabatic or WKB approximation. Equation~\eqref{eq:chieq} has a
solution of the form
 \begin{equation}
\chi_k(\eta) = \frac{1}{\sqrt{2 W_k(\eta)}} \; \exp\left(-i \int^\eta W_k(\eta') \, d\eta' \right)\,,
 \label{eq:WKB}
 \end{equation}
where $W_k(\eta)$ is a solution of
 \begin{equation}
 [W_k(\eta)]^2 = k^2 + V(\eta) -\frac{1}{2}\, \left(\frac{W''_k}{W_k} - \frac{3 (W'_k)^2}{2 W_k^2} \right)\,
 \label{eq:Wk}
 \end{equation}
where $W'_k = dW_k/d\eta$. When the expansion rate is sufficiently slow that the derivative terms in the above
equation may be ignored, we have the familiar WKB approximation 
 \begin{equation}
\chi_k(\eta) \approx \frac{1}{\sqrt{2 \omega_k(\eta)}} \; \exp\left(-i \int^\eta \omega_k(\eta') \, d\eta' \right)\,,
 \label{eq:WKB0}
 \end{equation}
where 
\begin{equation}
  \omega_k(\eta)  = \sqrt{k^2 + V(\eta)} \,.
  \label{eq:omega0}
 \end{equation}
This approximation is valid when both $(a')^2$ and $|a''(\eta)|$ are small compared to $[\omega_k(\eta)]^2$. Note that this can be viewed
as either a slow expansion, or a high frequency approximation. 

Note that Eq.~\eqref{eq:WKB0} is a positive frequency solution, with a slowly varying frequency. It may be taken to define the adiabatic vacuum
state, and hence adiabatic particle number. So long as the WKB approximation holds, this form for $\chi_k(\eta)$ remains a positive frequency mode,
and hence no particle creation occurs and particle number remains constant. Thus, the WKB solutions may be used for an approximate local
definition of particle number. This approach has been used by several authors~\cite{Woodhouse76,Berger78,Azuma83,Winitzki05,Parker12,Agullo15}.

Another attempt to give a local definition of particle number is the Hamiltonian diagonalization approach of Grib and Mamaev~\cite{Grib69}. Here one
defines a Hamiltonian, diagonalizes it by a Bogolubov tranformation, and defines particles in the new Fock space after diagonalization. The physical
motivation for this definition is not clear, and it has been critically discussed by Fulling~\cite{Fulling79}, who argues that the choice of Hamiltonian
is ambiguous. However, Fulling also notes that in some cases Hamiltonian diagonalization is essentially equivalent to the WKB approach. 

Several other criteria for defining particular classes of quantum states have been discussed in Refs.~\cite{Dray83,Degner,AP-Lim03,Raine75}.

\section{Quantum State of the Created Particles, Decoherence, and Entropy}
\label{sec:state-entropy}

\subsection{The Quantum State}
\label{sec:state}

Let us return to the Bogolubov transformation, Eq.~\eqref{eq:Bogo1} for particle creation in a spatially flat universe. In
this case, we have 
\begin{equation}
a_{\bf k}= \alpha _{k}^* b_{\bf k}     - \beta _{k}^*  b_{\bf -k}^\dagger \,,
                                      \label{eq:Bogo-flat}
\end{equation}
where $a_{\bf k}$ is the annihilation operator for a particle in mode ${\bf k}$   in the in-region, and $b_{\bf k}$ is the
corresponding operator in the out region. Recall that the quantum state of the system, when no particles are initially
present, is the in-vacuum $|0\rangle_{in}$, for which $a_{\bf k} |0\rangle_{in}=0$. The effect of the expansion of the universe
is to create pairs of particles in the out-region, where one member of each pair is in mode ${\bf k}$, and the other in
mode ${\bf - k}$. For a moment we ignore all other modes, and write the quantum state in the out-region as
 \begin{equation}
 |0\rangle_{in}= \sum_{n=0}^\infty c_n\, |n,n \rangle \, ,
 \end{equation}
where  $|m,n\rangle$  a number eigenstate in the out-region with $m$ particles in mode ${\bf k}$, as well as $n$ particles in mode ${\bf -k}$.
Thus
\begin{equation}
  b_{\bf k}^\dagger b_{\bf k} |m,n\rangle =   m\,   |m,n\rangle \,,  \quad b_{\bf -k}^\dagger b_{\bf -k} |m,n\rangle  = n \,  |m,n\rangle \,.
 \end{equation}
Now we have
 \begin{equation}
 a_{\bf k}|0\rangle_{in}=  \sum_{n=0}^\infty    c_n \,[  \alpha _{k}^* \sqrt{n}\,  |n-1,n \rangle -  \beta _{k}^*\, \sqrt{n+1}\, |n,n+1 \rangle 
 = \sum_{n=0}^\infty   ( \alpha _{k}^* \sqrt{n+1}\, c_{n+1} - \beta _{k}^*\, \sqrt{n+1}\, c_n)  |n,n+1 \rangle  = 0\, ,
 \end{equation}
which leads to the recurrence relation
 \begin{equation}
 c_{n} = \frac{\beta _{k}^*}{\alpha _{k}^*} \, c_{n-1} = \left(\frac{\beta _{k}^*}{\alpha _{k}^*}  \right)^n \, c_0 \,.
 \end{equation}
The normalization of the state $ |0\rangle_{in}$ yields
\begin{equation}
 \sum_{n=0}^\infty |c_n|^2 = |c_0|^2 \,  \sum_{n=0}^\infty \left|  \frac{\beta _{k}}{\alpha _{k}} \right|^{2n} = \frac{ |c_0|^2 }{1- |\beta _{k}/\alpha _{k}|^2} =  |c_0|^2 \,|\alpha _{k}|^2  =1  \,.,
  \end{equation}
where the second from last step follows from Eq.~\eqref{eq:alphabeta2}. 

This argument determines the state $|0\rangle_{in}$ in the out-Fock space up to an overall phase, so we have
 \begin{equation}
 |0\rangle_{in} = c_0  \sum_{n=0}^\infty \left(\frac{\beta _{k}^*}{\alpha _{k}^*}  \right)^n  \,  |n,n \rangle \,,
 \end{equation}
where $|c_0| = 1/|\alpha _{k}|$. From this expression, we see that the state of the system in the out region is a superposition of all possible number of pairs of particles, where
within each pair, the two particles have opposite momenta. Furthermore, the probability of finding $n$ pairs in this state is
 \begin{equation}
 P_n = |c_n|^2 = |\alpha _{k}|^{-2} \,  \left|  \frac{\beta _{k}}{\alpha _{k}} \right|^{2n} \,.
 \end{equation}

So far , we have considered only one value of ${\bf k}$, but particles are expected to be created into all modes for which $\beta _{k} \not= 0$. We may generalize the above
discussion to the case of multiple values of ${\bf k}$. The result may be written as
 \begin{equation}
 |0\rangle_{in} = c_0 \sum_{\{n_{\bf k}\}}  \prod_{\bf k}  \left(\frac{\beta _{k}^*}{\alpha _{k}^*}  \right)^{n_{\bf k}}   \; |\{n_{\bf k}\} \rangle \,.
 \label{eq:state}
 \end{equation}
Here $\{n_{\bf k}\}$ denote a set of occupation numbers; for each mode ${\bf k}$ there is non-negative integer $n_{\bf k}$ which gives the number of particles in mode ${\bf k}$
and the number of particle in mode ${-\bf k}$. In our enumeration, we may require $k_z \geq 0$ to avoid over counting. The overall constant $c_0$ satisfies
 \begin{equation}
 |c_0| = \left(\prod_{\bf k} |\alpha_k| \right)^{-1}\,.
 \end{equation}
The quantum state of the created particles is  a multi-mode squeezed vacuum state.

\subsection{Correlations, Entropy and Decoherence}
\label{sec:entropy}

The quantum state, Eq.~\eqref{eq:state}, is a highly entangled state with numerous quantum correlations. The two-particle sector of this state describes
pairs of particles with equal and opposite momenta. In effect, each created  particle is entangled with a partner moving in the opposite direction. In
addition, there are sectors of the state with all possible numbers of pairs, which are in turn correlated with one another. So long as the state of the system
is a pure quantum state, then its entropy is zero. However, one can define a formal entanglement entropy, or von Neumann entropy, for such a state. This is
discussed by Lin, {\it et al}~\cite{Lin2010}, who show that the von Neumann entropy  for the state Eq.~\eqref{eq:state} may be written as
 \begin{equation}
 S_{vN} = \sum_{\bf k} [(n_{\bf k}  +1)\, \ln(n_{\bf k}  +1) - n_{\bf k} \, \ln n_{\bf k} ]\,,
 \label{eq:SvN}
 \end{equation}
where $n_{\bf k} = |\beta _{k}|^2$ is the mean number of particle in mode $ {\bf k}$.  As before, we are assume quantization in a box of coordinate volume
${\cal V}$, and in the large volume limit, the sum on discrete modes is replaced by an integral, as in  Eq.~\eqref{eq:large-V}. This shows that $S_{vN}$ is
an extensive quantity proportional to  ${\cal V}$, as expected. 

The physical meaning of entanglement entropy is that it may be converted into thermodynamic entropy by decoherence, which is the loss of phase information
in the quantum state through interaction with an environment. The details of this process need not concern us, so long as the mean number of particles per
mode is approximately constant. Consider the case of a massless scalar field, and assume that after decoherence,  $n_{\bf k}$ is replaced by a Planck factor:
 \begin{equation}
 n_{\bf k} \rightarrow \frac{1}{{\rm e}^{\omega/(k_B \,T) } -1}\,,
 \end{equation}
where $\omega = |{\bf k}|$ and $k_B$ is Boltzmann's constant. Then  $S_{vN}$ becomes the usual thermodynamic entropy $S_T$ at temperature $T$,
 \begin{equation}
 S_{vN} = S_T = \frac{2 \pi^2}{45} \, {\cal V}\, (k_B \,T)^3\,.
 \end{equation}
This is the expected entropy of massless scalars at temperature $T$ and is
one-half of the familiar expression for the entropy of electromagnetic blackbody radiation.

\section{Applications to Cosmology}
\label{sec:cosmology}

\subsection{Inflation}
\label{sec:inflation}

\subsubsection{Essentials of Inflation}

Inflationary models postulate a period of accelerating expansion in the early universe~\cite{Starobinsky80,Guth81}. 
These models successfully solve some previous cosmological puzzles, the horizon
and flatness problems,  and predict a nearly scale invariant spectrum of initial perturbations for the post-inflationary universe as well as a universe which is very close
to the spatially flat model of Eq.~\eqref{eq:metric}. For a recent review, see Ref.~\cite{Vazquez20}. 
During inflation, the spacetime  metric is approximately  that of deSitter space, given by
 \begin{equation}
 a(t) = {\rm e}^{H\,t} =a(\eta) = -\frac{1}{H \, \eta}\, , \qquad \eta < 0\,
 \label{eq:deS}
 \end{equation}
where $H$ is the Hubble parameter, a constant with dimensions of inverse time. deSitter space is a solution of Einstein's equations with a positive cosmological constant,
or, equivalently, a matter stress tensor which is proportion to the metric tensor and an energy density of
 \begin{equation}
\rho_I =  \frac{3\, H^2}{8 \pi}\,.
\label{eq:vac-energy}
 \end{equation}
There are many versions of inflationary cosmology, but one of the simplest assumes a classical scalar field (the 'inflaton") with an effective potential which has a flat region,
where the first derivative of the potential with respect to the field is small.
This field obeys an equation analogous to that of a classical particle in this potential. When the field is in the flat region, its energy density is approximately constant and equal
to $\rho_I$. This is the inflationary period, which is assumed to end at a finite value of $t=t_R$ or of $\eta= \eta_R$. As inflation ends, the energy in the scalar field is converted into ordinary 
matter particles. If these particle are relativistic, then the subsequent epoch of the universe is a radiation dominated expansion, where
 \begin{equation}
 a(t) \propto (t-t_R)^{1/2}\, , \quad {\rm or} \quad a(\eta) \propto \eta - \eta_R \,.
 \label{eq:rad-dom}
 \end{equation}

Apart from  a successful resolution of the horizon and flatness problem, one of the great successes of inflationary cosmology is the prediction of a nearly scale invariant spectrum
on initial density perturbations, which has been confirmed by several observations, especially the temperature fluctuations of the cosmic microwave background. These density
perturbations arise from quantum fluctuations of the inflaton field during inflation~\cite{CM81,GP82,H82,S82,BST83}. For a review, see Ref.~\cite{Brandenberger85}.
 This effect may be visualized as follows: the inflaton is a nearly classical field, but has small
vacuum fluctuations, which cause some spatial regions to evolve slightly faster than nearby regions. The regions which evolve faster transition from the deSitter phase to a 
radiation dominated phase somewhat sooner and begin redshifting sooner. At later times, these regions have a lower temperature and smaller energy density compared to regions
which ended inflation later. In addition to these density, or scalar, perturbations, inflation also predicts tensor  perturbations in the form of primordial gravity waves which will be
discussed in Sec.~\ref{sec:graviton-inflation}.

 The creation of matter at the end of inflation ('reheating") is usually assumed to occur due to self-coupling of the inflaton field and its coupling to other matter fields. This reheating period
 necessarily requires a finite interval of time, during which the spacetime metric makes a smooth transition from that of deSitter space to that of a radiation dominated universe.
 For recent reviews of reheating, see Refs.~\cite{Allahverdi2010,Amin2015}. In addition
 to matter creation by field couplings, there should be at least some particle creation by the mechanism described above in Sec.~\ref{sec:particle-creation}, which is the principle topic of
 this review. This is sometimes called gravitational particle creation to distinguish it from  creation by field couplings.
 
 \subsubsection{Initial Conditions: The Bunch-Davies Vacuum}
 \label{sec:Bunch-Davies}
 
  One of the requirements for well-defined quantum particle creation are clear definitions for the in and out vacua. In our discussion in Sec.~\ref{sec:particle-creation}, both the initial and
  final regions were taken to be flat spacetime regions. In the context of cosmology, the final approximately flat region is not a problem, but the initial one is problematic. Fortunately,
  inflation seems to offer a natural resolution of this difficulty. An essential feature of a long period of inflationary expansion is that it effectively erases the memory of whatever preceeded
  it through redshifting of any preexisting matter. Regardless of the quantum state of the matter fields at the beginning of inflation, the effective state rapidly approaches the vacuum state
  as inflation proceeds. For massless conformal fields there is no ambiguity in the definition of the vacuum in a conformally flat spacetime, such as that of Eq.~\eqref{eq:metric}, and the
  natural definition is called the conformal vacuum. For massive or other nonconformal  fields, there is generally a choice of vacuum in deSitter space which is deSitter invariant, and
  respects the spacetime symmetries of deSitter space. This is often called the Bunch-Davies vacuum~\cite{BD78}, and is the state which any initial state will asymptotically approach
   after an extended period of inflation. 
  There is a subtlety in the case of the massless, minimally coupled scalar field in that the two point function suffers from an infrared divergence, so in a strict sense, the
  Bunch-Davies vacuum does not exist for this field. However, there are a family of quantum states which are infrared finite and any initial state of the massless, minimally coupled scalar field 
  will approach one of these states~\cite{FV86}. These states are also close to the Bunch-Davies state in the sense that expectation values of quantities such as the stress tensor, which are free of
  infrared divergences, due to the presence of space or time derivatives, are the same in all of these states as in the Bunch-Davies state.
  Thus it is reasonable to take the Bunch-Davies vacuum as the in-vacuum for discussions of quantum particle creation during the transition from inflationary expansion to the 
  radiation dominated universe. Note, however, that particles in very high frequency modes in the initial state could have an effect if these modes are not sufficiently redshifted
  by the inflationary expansion. This possibility was discussed by Agullo and Parker~\cite{Agullo11}, who consider the effect of stimulated emission produced by such particles.

\subsubsection{Estimates of Gravitational Particle Creation during Reheating}
\label{sec:reheating}

Here we wish to estimate the magnitude of the number and energy densities of particles which are created by the transition from deSitter expansion to a radiation dominated epoch. We follow
the treatment in Ref.~\cite{F87}, using the perturbation method of Sec.~\ref{sec:Pert}. Consider the case of massless, non-conformally coupled scalar particles, and set $a=a_f=1$ at the end
of reheating, $\eta = \eta_R$. The scalar curvature vanishes in a spatially flat radiation dominated universe, so take $R=0$ for $\eta > \eta_R$. The particle creation effectively ceases at
$\eta = \eta_R$, and we can estimate the number density of created particles in the subsequent phase from Eq.~\eqref{eq:Npert} to be 
 \begin{equation}
n \approx  \frac{(\xi - \frac{1}{6})^2}{16\pi a^3} \int_{-\infty}^{\eta_R} a^4(\eta)\,R^2\,d\eta = \frac{(\xi - \frac{1}{6})^2\; H^3}{12 \pi\, a^3}\,,
\label{eq:n-inf}
 \end{equation}
where we have used the fact that $R = 12 H^2$ during inflation, and the factor of $a^{-3}$ describes the dilution of the particle  number density by the cosmological expansion after their
creation. If we assume that the transition occurs on a comoving time scale of $\Delta t \alt 1/H$, then it is shown in Ref.~\cite{F87} that Eq.~\eqref{eq:rhopert} implies that the energy
density of the created particle is
 \begin{equation}
\rho \approx  \frac{(\xi - \frac{1}{6})^2\, H^4}{8\pi^2\, a^4} \; \ln[1/(H \Delta t)]\,. 
\label{eq:rho-inf}
 \end{equation}
 This implies that the mean energy of the created particles is of the order of $H \,\ln[1/(H \Delta t)]$. Note that, unlike the number density, the energy density and the mean energy per
 particle diverge as $\Delta t \rightarrow 0$. This reflect a general feature that very rapid changes in spacetime geometry can lead to violent particle creation.

The inflaton energy density, $\rho_I$, which drives the expansion, is related to
$H$ by the Einstein equation:
\begin{equation}
H^2 = {\frac{8\pi\rho_I}{3\sqrt{\rho_{Pl}}}} \, ,  
\end{equation}
where $\rho_{Pl} \approx \bigl( 10^{19} GeV\bigr)^4$ is the Planck density.
We can express our estimate for the energy density of the created particles
just after the end of inflation as
\begin{equation}
\rho \approx (1-6\xi)^2 {\frac{\rho_I^2}{\rho_{Pl}}}.
\label{eq:rho-inf2}
\end{equation}
If, for example, we were to take $\rho_I \approx \bigl( 10^{15}\, GeV\bigr)^4$,
 then we obtain the estimate 
\begin{equation}
\rho \approx (1-6\xi)^2 \left( 10^{11}\, GeV\right)^4.
\end{equation}
This energy density is much less than $\rho_I$, and would hence be negligible 
if there is efficient reheating. However, if other reheating mechanisms are
not efficient, then particle creation by the gravitational field could
play a significant role in cosmological evolution. 

However, a very weak coupling between the inflaton and other fields has some phenomological advantages,
as it facilitates  a very long period of slow roll needed to have enough inflationary expansion to solve the horizon
and flatness problems. Peebles and Vilenkin~\cite{PV99} have discussed this issue and constructed a plausible
inflationary model in which gravitational particle creation plays a key role. Some  more recent work on the
role of gravitational particle creation in various reheating models has been given, for example, in 
Refs.~\cite{Koutsoumbas13,Moghaddam2017,Haro19,Hashiba19,Hashiba2019,Lankinen2020}.
Dissipation and entropy production have been discussed in Refs.~\cite{Kandrup88a,Kandrup88b,Calzetta89}.

\subsubsection{DeSitter Particle Creation?}
\label{sec:DeSitter}

There have been many papers written claiming that deSitter space itself is unstable to runaway quantum particle creation.
See, for example, Mottola~\cite{Mottola85}. These claims are controversial, and have been disputed by Marolf and
Morrison~\cite{MM10,MM11}, who give arguments in favor of the quantum stability of deSitter space. The model described
in the previous subsection, which leads to Eq.~\eqref{eq:rho-inf}, seems to support the quantum stability, at least so
far as quantum particle creation is concerned. Here the energy density of created particles depends upon the transition
time from de Sitter to the radiation dominated universe, but does not depend upon the duration of the de Sitter phase. 
If runaway particle creation were occurring, one would expect this energy density to grow with increasing duration of
inflation. This appears to be a situation where the questions of initial conditions and definition of particle number,
discussed in Sec.~\ref{sec:initial}, play an important role.

\subsubsection{Dark Matter and Baryogeneis from Gravitational Particle Creation}
\label{sec:dark-matter}

One of the mysteries of modern cosmology is the nature of the dark matter, which  composes about 25\% of the mass density of
the Universe. This is in contrast to only about 5\% due to stars and other luminous matter. The dark matter reveals its presence
only through its gravitational effects. 
Furthermore, the fraction of dark matter appears to grow as one probes larger length scales. For example, the dynamics of galaxy
clusters reveal a larger dark matter to luminous matter ratio than do the rotation curves of individual spiral galaxies.

One plausible explanation of the dark matter is that it consists of weakly interacting, massive particles (WIMPS). The difficulty in
directly detecting dark matter particles requires that they interact only very weakly with ordinary matter particles. However, this makes
formation of dark matter in the early universe more difficult. One possibility is that the dark matter couples only through the gravitational
interaction, which is vastly weaker than the other known forces of nature. In this case, gravitational particle creation becomes a
promising model for the origin of the dark matter, and has been explored by many 
authors~\cite{Chung1998,Chung2001,Chung2012,Ema18,Ema2019,Li2019,Cembranos20,Ahmed20,Herring2020a,Herring2020b}.
Several  models have been explored,
including ones where the present dark matter particles were directly created by the expansion of the Universe at the end of inflation, as described 
above. Another possibility is that very massive particles were initially created, and subsequently decayed into lighter 
particles~\cite{Chung1998,Chung2001,Li2019}. Massive particles created at the end of inflation have also been proposed
as  models for gamma rays~\cite{Dannehold80} and ultra high energy cosmic rays~\cite{Kuzmin1999,Kuzmin99,Dick06}. 

Gravitational particle creation could also play a role in baryogenesis, the creation of a net baryon number in the universe. This has been discussed
in Refs.~\cite{Davoudiasl04,Lima16}.

\subsubsection{Graviton Creation in Inflation and Tensor Perturbations}
\label{sec:graviton-inflation}

The transition from deSitter expansion to a subsequent radiation dominated phase will not only create matter particles,
but also gravitons, which can later appear as primordial gravity waves which will contribute to the net energy density
of the universe, but also lead to anisotropic, tensor perturbations of the cosmic microwave background (CMB). This possibility
was first suggested by Starobinsky~\cite{S79}, who used a  somewhat heuristic procedure of matching  a graviton mode
function in deSitter space to one in the radiation dominated phase. In the transverse tracefree gauge, this amounts to
matching two different solutions of Eq.~\eqref{eq:chieq} with different choices of $a(\eta)$ but $m=0$ and $\xi =0$.
The resulting mode in the radiation dominated phase  is interpreted as describing a classical, stochastic gravity wave.
A similar procedure was used in Refs.~\cite{RSV82,FP83,AW84a,AW84b}, where estimates of the resulting anisotropy
of the CMB were given. Later Abbott and Harari~\cite{AH86} and Allen~\cite{A88} gave more rigorous treatments using
the Bogolubov coefficient approach.

After their creation, the gravitons are in a squeezed vacuum state of the form of that found in Sec.~\ref{sec:state}. At
this point, they are far from describing a classical gravity wave, as the expectation values of the metric and curvature
perturbation vanish in such a state. In order to become a classical wave, the gravitons need to decohere into a state
closer to a coherent state.
Once this has occurred, the resulting classical waves can produce anisotropies in the temperature and polarization of the
CMB which are potentially  observable, but have not yet been detected.

In many models of inflation, the inflaton field is expected to oscillate around a minimum value as reheating proceeds.
These field oscillations can drive metric oscillations, which can in turn lead to gravitational creation of gravitons and
of other particles. This is distinct from particle creation by direct coupling of matter fields to the istanton, Gravitational
particle creation by metric oscillations at the end of inflation was discussed by Vilenkin~\cite{Vilenkin:1985} in the context of the
Starobinsky model~\cite{Starobinsky80}, and by Suen and Anderson~\cite{SA87} in other  versions of inflation.
Graviton creation was discussed in Ref.~\cite{Ema:2015}  as a mechanism for inflaton decay. 
Several other authors have discussed various aspects of gravitons created in inflationary 
models~\cite{Yainik90,Grishchuk-Sidorov90,Maia93,Henriques94,Mendes99,Allen2000,Henriques04,Giovannini20}.

\subsubsection{Effects of Anisotropy and Inhomogeneity}
\label{sec:aniso}

It has been noted by several authors that anisotropic expansion can lead to enhanced particle creation. Before the inflationary model
was proposed as a solution to the remarkable observed isotropy of the universe, particle creation was investigated as a damping
mechanism for anisotropic expansion~\cite{ZS,Hu73,Hu74,Berger75,Hu78}.  The effects of small anisotropy may be studied in the
perturbative method discussed in Sect.~\ref{sec:Pert}. Anisotropy results in a nonzero Weyl tensor, $C_{\alpha\beta\mu\nu}$.
For the case of massless scalar particles, Eq,~\eqref{eq:Npert} for the mean number density of created particles, now
becomes~\cite{ZS,BD2}
\begin{equation}
n= \frac{1}{16\pi a_f^3} \int_{-\infty}^\infty  a^4(\eta)\,
 \left[ \frac{1}{60}  C_{\alpha\beta\mu\nu} C^{\alpha\beta\mu\nu}  +  \left(\xi - \frac{1}{6}\right)^2\, R^2(\eta)\right]\,d\eta \,.
                        \label{eq:Npert2}
\end{equation}
Thus the anisotropy produces a contribution to the average particle creation rate which is proportional to the square of the
Weyl tensor. Note that particles are created even in the conformal limit, $\xi = 1/6$. More generally, massless conformal
particles, including photons and fermions~\cite{Bhoonah19}, can be created in an anisotropic universe. Anisotropy can also enhance entanglement
entropy, as was recently discussed by Pierini\,{\it et al}~\cite{Pierini19}.

Although inflation removes the immediate motivation for anisotropy damping mechanisms, it leaves other questions unresolved.
One of these is the issue of the set of initial conditions which can give rise to inflation. This has been discussed by several
authors, including Calzetta~\cite{Calzetta91}, who argues that particle creation can enlarge this set of initial conditions.

The treatments of anisotropy summarized above still assume a homogeneous universe. However, they may be generalized
to include the effects of small inhomogeneities. This was done by   Frieman~\cite{Frieman89}  and by Cespedes and
Verdaguer~\cite{Cespedes90}.

\subsection{Late Time Particle Creation}
\label{sec:late-time}

After inflation has ended, and the universe has entered a radiation or a matter dominated epoch, gravitational particle
creation would normally be expected to become very small, as the spacetime curvature has become relatively small,
and the expansion time correspondingly large. However, there are some possible exceptions, including emergent 
cosmology models. These models postulate a universe which undergoes a very long, and possibly infinite, period of oscillations
before emerging into an inflationary universe. Emergent models have been discussed by Bag, {\t et al}~\cite{Bag:2014},
who show that quantum creation of gravitons during the oscillatory phase places strong constraints on such models.

Another possible source of late time metric oscillation arise from effective actions for gravity which contain terms quadratic
in the curvature. Such terms can arise either in modified classical theories of gravity or from one loop quantum corrections,
and can produce rapid metric oscillations around an expanding background spacetime~\cite{HorowitzWald:1978}.
These oscillations can produce gravitons, as was discussed in Refs.~\cite{SF16,Ema2016}. The scale factor, including the osciilations,
can be taken to be
\begin{equation}
a(t)= a_B(t)\, [1 + A_0 \cos(\omega_0 \, t)] \, , 
\label{scalefactor}
\end{equation}
where $A_0$ and $\omega_0$ are the amplitude and angular frequency of the oscillations, and $ a_B(t)$ is the background
scale factor. The graviton creation rate may be calculated using the perturbative approach described in Sec.~\ref{sec:Pert}.
On time scales short compared to the background expansion time, both the graviton number density and energy density seen
by a local comoving observer grow linearly in time:
\begin{equation}
n_{g}\ = \frac{A_0^2\,\omega_0^4t}{16\pi} \, ,
\label{eq:gravitonnumberdensity}
\end{equation} 
and
\begin{equation}
\rho_{g} = \frac{A_0^2\,\omega_0^5\,t}{32\pi} \, .
\label{eq:gravitonenergydensity}
\end{equation}

Comparison of these two expressions reveals that the mean energy per graviton is $\omega_0/2$ This result may be understood from
local energy conservation. In general, energy is not conserved in a time-dependent background spacetime, but here we may take the
rate of variation of $a_B(t)$ to be small compare to the oscillatory factor. In this case, we effectively have oscillations around  an average
background of flat spacetime, and the time averaged energy is conserved. The metric oscillation may be viewed as a classical gravity
wave at angular frequency $\omega_0$, or equivalently, as a coherent state of gravitons with energy $\omega_0$. This energy is converted
into a pair of quantum created gravitons, each with energy $\omega_0/2$. This effect is similar to that of parametric nonlinear down conversion
in nonlinear optics. (See, for example, Ref.~\cite{Couteau}.) 

The energy density of the created gravitons contributes to net energy density which determines the overall expansion rate of the universe,
and may be constrained by cosmological observations. This in turn places constraints on the parameters $A_0$ and $\omega_0$ and
on the underlying modification of gravity theory.

Another context in which quantum particle creation could become important in the late universe would if the effective equation of state of
the matter in the universe were of the form $p = w\, \rho$, where $w < -1$, sometime called phantom matter. The equation of state causes
such rapid expansion that curvature singularity arises, the ``Big Rip". Nunes and Pavon~\cite{Nunes15} have suggested that  quantum
 particle creation could give rise to such an equation of state. However,  Dimopoulos~\cite{Dimopoulos18} argues that the backreaction
 from particle creation could have the effect of increasing the value of $w$, and preventing the Big Rip.

\section{Backreaction}
\label{sec:backreact}

In this section, we briefly address the backreaction of particle creation on the background spacetime geometry. Here we assume a semiclassical version of general
relativity in which the expectation value of the stress tensor operator of the quantum field acts as part of the source of the gravitational field. Thus Einstein equation
in this theory takes the form
 \begin{equation}
 G_{\mu \nu} = 8 \pi G\, (T^{cl}_{\mu \nu} + \langle \hat{T}_{\mu \nu}  \rangle)\,,
 \label{eq:SC-Einstein}
 \end{equation}
where $G_{\mu \nu}$ is the Einstein tensor of the classical background spacetime, $G$ is Newton's constant, $T^{cl}_{\mu \nu}$ is the stress tensor of any
classical matter present, $\hat{T}_{\mu \nu}$ is the quantum stress tensor operator, and $\langle \hat{T}_{\mu \nu}  \rangle$ is its suitably defined expectation value
in some quantum state. Here the phrase ``suitably defined'' hides a great deal of technical complexity, as the formal expectation value suffers from ultraviolet
divergences which need to be removed by a combination or regularization and renormalization. Here we give only a quick summary, and refer the reader
to, for example Refs.~\cite{BD,PT}, and the papers cited therein for more details.

Regularization is a formal prescription for removing the divergences, but leaving a result which depends upon one or more undetermined parameters. The conceptually
cleanest regularization methods are covariant ones which respect the general covariance of the theory, such as dimensional regularization. In this case, the 
regularized expectation value is finite, but contains four terms which would diverge if the regulator were removed. These terms are proportion to the spacetime
metric tensor $g_{\mu \nu}$,   the Einstein tensor, and two tensors which are quadratic in the curvature, $H^{(1)}_{\mu \nu}$ and $H^{(2)}_{\mu \nu}$. The latter two
tensors may be taken to be the functional derivatives of $R^2$ and of $R_{\mu \nu}R^{\mu \nu}$, respectively, with respect to the metric tensor.  The regulator
dependence may be absorbed by renormalization, which is a redefinition of the cosmological constant, Newton's constant, and two constants which are introduced
as coefficients of $R^2$ and of $R_{\mu \nu}R^{\mu \nu}$ in the gravitational action. In writing Eq.~\eqref{eq:SC-Einstein}, we have assumed that the renormalized
values of all of these constants have been set to zero, apart from $G$, whose renormalized value is that measured by the Cavendish experiment.. 
However, the renormalized expectation value can still contain finite terms proportional to each of the four tensors, $g_{\mu \nu}$, $G_{\mu \nu}$, $H^{(1)}_{\mu \nu}$,
and  $H^{(2)}_{\mu \nu}$. The latter two terms can cause particular problems because they involve fourth derivatives of the metric tensor and lead to a fourth-order
Einstein equation, which can suffer from instabilities or oscillations as discussed in Sec.~\ref{sec:late-time}. One resolution to this problem has been discussed by
Simon~\cite{Simon90}, who argues that higher order theories can be viewed as resulting from approximations to a nonlocal theory, and may be treated in perturbation
theory with the imposition of constraints which eliminate the unstable solutions.

More generally, it is not possible to uniquely distinguish contributions to  $\langle \hat{T}_{\mu \nu}  \rangle$ coming from created particles from those due to vacuum
energy effects. This is related to the intrinsic ambiguity in defining particle number in a dynamic spacetime. However, we may make some order of magnitude
estimates of the combined effects of particles and vacuum energy. Let $\ell$ be the characteristic length scale associated with the spacetime curvature. In the case of
an expanding spatially flat universe, such as that described by Eq.~\eqref{eq:BD-scale}, $\ell$ is the characteristic expansion time scale, $1/\rho$. The Einstein and
Ricci tensors are of order $1/\ell^2$,
 \begin{equation}
 G_{\mu \nu} \approx R_{\mu \nu} \approx O\left(\frac{1}{\ell^2}\right) \,,
 \end{equation}
 while  $\langle \hat{T}_{\mu \nu}  \rangle$  is typically of order $1/\ell^4$, so 
 \begin{equation}
 G \, \langle \hat{T}_{\mu \nu}  \rangle \approx O\left(\frac{\ell_P^2}{\ell^4}\right) \,,
 \end{equation}
where $\ell_P$ is the Planck length. Thus, we expect the backreaction from quantum effects on the background geometry to be small if $\ell \gg  \ell_P$.
However, as noted in Sect.~\ref{sec:reheating} , even though the energy density of particles created at the end of inflation may be small compared to the vacuum energy during
inflation (See Eq.~\ref{eq:rho-inf2}.), the effects of the created particles can still be significant.
 
 Although $\langle \hat{T}_{\mu \nu}  \rangle$  can contain contributions both from created particles and from vacuum energy which are hard to distinguish, there are cases
 where this distinction may be made. One case is an out region after particle creation has ceased, as in  Eq.~\eqref{eq:BD-scale} when $\eta \gg \rho$. Here spacetime is essentially
 flat so there is no vacuum energy effect, but there can be a nonzero energy density of created particles. Another case is that of a conformally invariant field.. Here there is no 
 particle creation, but $\langle \hat{T}_{\mu \nu}  \rangle \not= 0$ during the expansion due to vacuum energy. 

In general, covariant regularization and renormalization of the stress  tensor in an expanding universe is very difficult to perform explicitly. An alternative approach is
the adiabatic regularization method of  Parker and Fulling~\cite{PF74}. This is based upon a generalization of the WKB method given by Chakraborty~\cite{Chakraborty}.
The key idea is to start with the solution of Eq.~\eqref{eq:chieq} given by  Eq.~\eqref{eq:WKB} and \eqref{eq:Wk}. To lowest order, one takes $W_k(\eta) \approx \omega(\eta)$,
given by Eq.~\eqref{eq:omega0}, but then successively iterates Eq.~\eqref{eq:Wk} to obtain a sequence of better approximations to $W_k(\eta)$. The leading terms in this sequence
will generate terms in the stress tensor proportional to the divergent contributions.  These terms are subtracted from the mode functions computed by some other means, such as
numerically. This leads to finite integrals over all modes for the stress tensor components.This method is especially well suited to numerical evaluation of the energy density, 
and has been used by several authors~\cite{AP87,SMPM99,ZK19}. 

The treatment of the backreaction on the background spacetime geometry requires a careful consideration of the choice of gauge. This is also an issue when discussing
the backreaction of classical perturbations, and has been discussed, for example, by Geshnizjani and Brandenberger~\cite{GB02}, who give a description in terms of a
modified local expansion rate.

\section{Analog Models}
\label{sec:analog}

In this section, we consider several effects which are analogous in some way to particle creation by gravitational fields.

\subsection{Schwinger Effect}
\label{sec:Schwinger}

An explicit example of quantum particle creation by a classical external field is the Schwinger effect~\cite{Schwinger},
the creation of electron-positron pairs by a constant electric field. The rate of pair creation per unit time per unit 
volume may be written as
\begin{equation}
R = \frac{\alpha^2}{\pi^2}\; E^2\; \sum_{n=1}^\infty n^{-2}\, \exp\left(-\frac{\pi\, n\, m^2}{e\, E} \right)\,.
\label{eq:R-Schwinger} 
 \end{equation}
Here $m$ and $e$ denote the mass and magnitude of the electric charge of the electron, respectively, $\alpha = e^2/(4 \pi)$ is the fine structure
constant, and $E$ is the magnitude of the electric field in Lorentz-Heaviside units . First we note that the pair creation is exponentially suppressed if $m^2 \gg e\, E$.
This is analogous to the suppression of the creation of massive particles by the cosmological expansion, as illustrated in Eq.~\eqref{eq:N-BD3},
for example.  In the electromagnetic case, this leads to a critical electric field, $E_{cr} = m^2/e \approx  1.3 \times 10^{18} V/m$, above which the spontaneous 
electron positron pair  creation rate becomes large. This type of particle creation may be derived by a Boglubov coefficient approach, of the type outlined in
Sect.~\ref{sec:Bogolubov}~\cite{Grib94}. The Schwinger effect has not yet been observed in the laboratory. However, there may be greater chance of
observing pair  creation in a time dependent laser field than in a static electric field~\cite{Dunne09}.

One can understand the above expression for $E_{cr}$ from a simple heuristic argument. The work on by an electric field $E$ moving a charge
$e$ through a distance $d$ is, from Newtonian mechanics, $W = e\, E\, d$. If we set this work to be of the order of the electron rest mass energy
and $d =\lambda_C = 1/m$ to be the electron Compton wavelength, then we obtain $E\approx E_{cr}$.

There is a key difference between the Schwinger formula, Eq.~\eqref{eq:R-Schwinger}, and the results for gravitational particle creation discussed
in Sect.~\ref{sec:specific}. The former depends upon a coupling constant, $e$, while the latter do not, presumably reflecting the universality of the
gravitational  interaction. Anderson, Mottola and Sanders~\cite{AM14,AMS18}   have argued that gravitational particle creation in global deSitter spacetime, 
as opposed to an inflationary universe, is analogous to the Schwinger effect. However, as discussed in Sec.~\ref{sec:DeSitter}, particle creation in  
deSitter spacetime does not seem to be well-defined.

\subsection{Particle Creation by Moving Mirrors}
\label{sec:mirror}

\begin{figure}[htbp]
\includegraphics[scale=0.25]{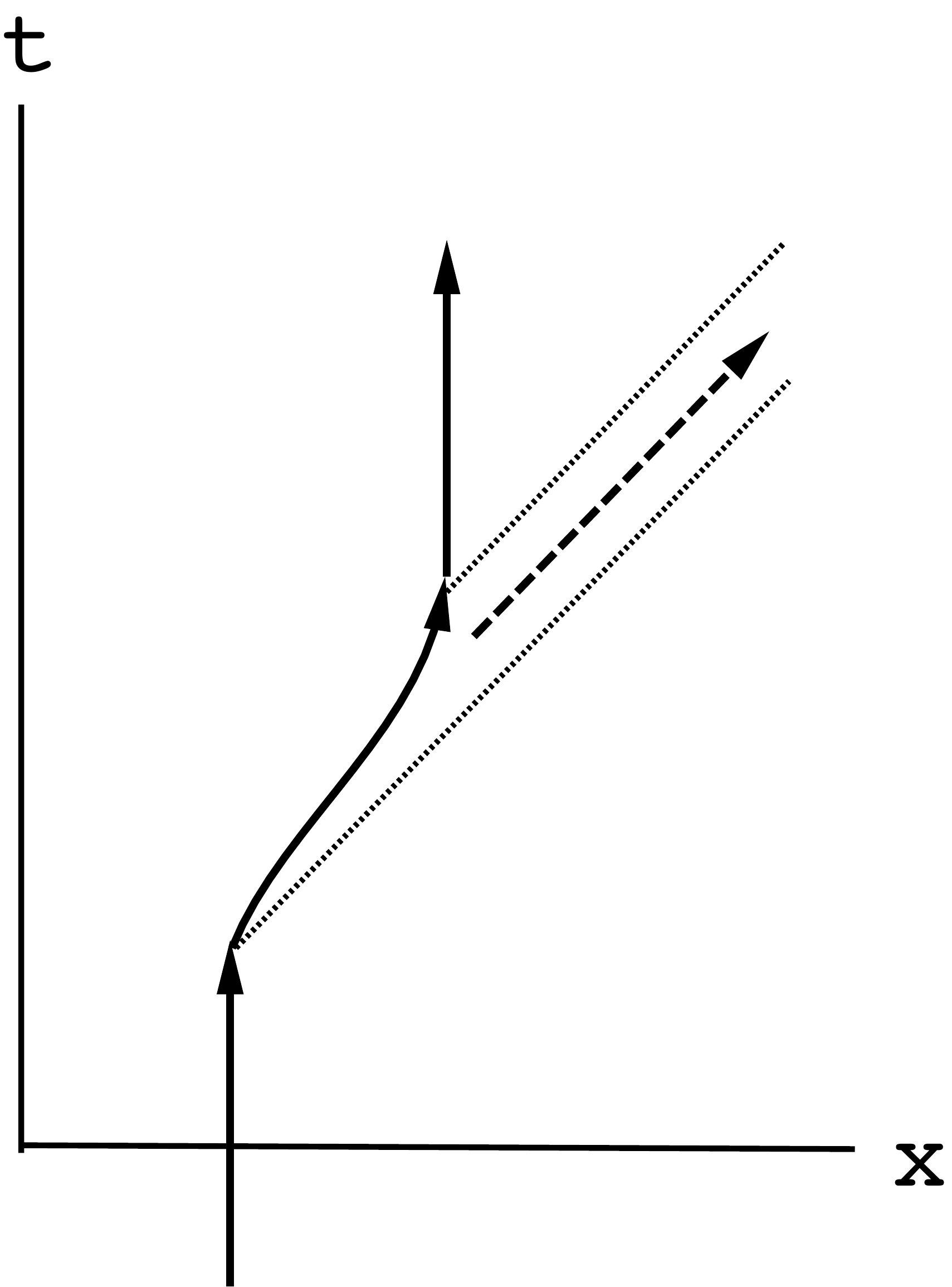}
\caption{A moving mirror in two-dimensional spacetime accelerates for a finite period of
time. The quantum radiation emitted to the right of the mirror propagates in
the spacetime region between the dotted lines. There is also radiation emitted
to the left which is not shown.}
\label{fig:mirror}
\end{figure}

    A simple example of quantum particle creation was given by Fulling and 
Davies \cite{FD76, DF77}. This consists of a moving mirror in two-dimensional
spacetime coupled to a massless scalar field, $\varphi$. The field is assumed
to satisfy a boundary condition on the worldline of the mirror, such as
$\varphi = 0$. For a given mirror trajectory, it is possible to construct
exact solutions of the wave equation which satisfy this boundary condition.
Let $v=t+x$ and $u=t-x$ be null coordinates which are constant upon null rays
moving to the left and to the right, respectively. A null ray of fixed $v$ 
which reflects off of the mirror becomes a ray of fixed $u$. The relation 
between the values of $v$ and of $u$ is a function determined by the mirror's
trajectory (See Figure~\ref{fig:mirror}.) Let
\begin{equation}
v = G(u) \,,
\end{equation}
or, equivalently, 
\begin{equation}
u = g(v) = G^{-1}(v) \,.
\end{equation}
The mode functions which satisfy the massless wave equation and which vanish
on the worldline of the mirror are
\begin{equation}
f_k(x) = \frac{1}{\sqrt{4 \pi \omega}}\, 
         \Bigl( e^{-i\omega v} - e^{-i\omega G(u)} \Bigr) \,.
\end{equation}
The incoming positive frequency wave, $e^{-i\omega v}$, is reflected from the
mirror and becomes an outgoing wave, $e^{-i\omega G(u)}$, which is a 
superposition of positive frequency  ($e^{-i\omega u}$) and negative
frequency ($e^{i\omega u}$) parts. 

   Fulling and Davies \cite{FD76} show that the flux of energy radiated to the 
right is
\begin{equation}
F(u) = \langle T^{xt} \rangle = 
\frac{1}{48 \pi} 
\biggl[ 3\biggl(\frac{G''}{G'}\biggr)^2 -2\biggl(\frac{G'''}{G'}\biggr)\biggr]
                     \,.  \label{eq:mirror}
\end{equation}
This flux may also be expressed in terms of the instantaneous mirror velocity
$v(t)$ as 
\begin{equation}
F = - \frac{(1-v^2)^{1/2}}{12 \pi (1-v)^2} \, \frac{d}{d t} \,
            \left[ \frac{\dot v}{(1-v^2)^{3/2}} \right] \, ,  \label{eq:mirror2}
\end{equation}
 In the nonrelativistic limit, 
\begin{equation}
F \approx - \frac{\ddot v}{12 \pi} \, .  \label{eq:mirror3}
\end{equation}
 Note that $F$ may be either positive or 
negative. In the latter case, one has an example of the negative energy in
quantum field theory.

   Unfortunately, the simple solution for the moving mirror radiation of a
massless field in two-dimensional spacetime depends upon the special conformal
properties of this case and does not generalize to massive fields or to 
four-dimensional spacetime. In the four-dimensional case, there are exact
solutions available for special trajectories \cite{CD77,FS79}, and approximate
solutions for general trajectories \cite{FV82a}, but no general, exact solutions.
However, the technique of mapping between ingoing and outgoing rays is crucial
in the derivation of particle creation by black holes~\cite{Hawking}.

Particle creation by moving mirrors is sometimes called the dynamical Casimir effect. An 
analog version of this effect has been observed in a metamaterial~\cite{LPHH13}. 
The possible effects of actual moving mirrors in an optical cavity have been discussed
in Ref.~\cite{Butera19}.

\subsection{Creation of Squeezed Light in Nonlinear Optics}
\label{sec:squeeze}

As noted above, the quantum state of the particles created by a time-dependent gravitational
field is a multi-mode squeezed vacuum state. In recent decades, it has become possible to
create squeezed states of photons in the laboratory using nonlinear optical materials~\cite{SHYMV85,WKHW86}. 
In this section, we will examine how this is done, and show that the basic physical process is essentially
the same as particle creation by an expanding universe or by a moving mirror.

First we review selected aspects of nonlinear optics. Recall that in SI units, the electric displacement
vector ${\bf D}$ in a material is related to the electric field  ${\bf E}$ and polarization (electric dipole
moment per unit volume) ${\bf P}$ by
 \begin{equation}
 {\bf D} = \epsilon_0 {\bf E}  + {\bf P}\,,
 \label{eq:D}
 \end{equation}
where $\epsilon_0$ is a constant. This relation holds in both isotropic and anisotropic materials.
In a nonlinear material, the Cartesian components of ${\bf P}$ become nonlinear function of those of ${\bf E}$,
and may be expanded in a power series of the form
\begin{equation}
P_i = \epsilon_0 \left(\chi_{ij}^{(1)} E_{j} + \chi_{ijk}^{(2)} E_{j}E_{k} 
+ \chi_{ijkl}^{(3)} E_{j}E_{k}E_{l} + \cdots \right) \,,
\label{eq:pol}
\end{equation}
where sums on repeated indices are understood. The coefficients in this expansion are tensors, and
are called susceptibilities. 

For simplicity, consider the case where the three vectors,  ${\bf D}$, ${\bf E}$, and ${\bf P}$ may be taken
to all be in the same direction, which we take to be the $z$-direction, so
 \begin{equation}
 D_i = \delta_{iz}\, D\,, \quad E_i = \delta_{iz}\,E\,, \quad   P_i = \delta_{iz}\, P\,.
 \end{equation}
Note that this does not require isotropy, but does require that certain components of the susceptibility 
tensors vanish, e.g. $\chi_{xz}^{(1)} =0$, ect. Now Eq.~\eqref{eq:pol} may be written as
\begin{equation}
P =  \epsilon_0 \left(\chi^{(1)} E + \chi^{(2)} E^2+ \chi^{(3)} E^3 + \cdots \right) \,,
\label{eq:pol2}
\end{equation}
where 
\begin{equation}
\chi^{(1)} = \chi_{zz}^{(1)} \,, \quad  \chi^{(2)} =  \chi_{zzz}^{(2)} \,, \quad
 \chi^{(3)} = \chi_{zzzz}^{(3)} \,.      \label{eq:chis}
\end{equation}
Here $\chi^{(1)}$ is the familiar linear susceptibility, $\chi^{(2)}$ is the second
order susceptibility, which can only be nonzero in a material whose unit cell lacks
spatial inversion symmetry, and  $\chi^{(3)}$ is the third order susceptibility.

Now let the electric field be a sum, 
\begin{equation}
 E= E_0 +E_1
 \end{equation}
of a background field $E_0$ and a perturbation $E_1$, where $|E_1| \ll |E_0|$,
and assume similar expressions for $D$ and $P$. If we substitute these forms into Eqs.~\eqref{eq:D}
and Eqs.~\eqref{eq:pol}, and expand to first order in the perturbation, we find
 \begin{equation}
 D_1 = \epsilon_{eff}\, E_1
 \end{equation}
 where 
  \begin{equation}
 \epsilon_{eff} = \epsilon_0\, [(1+ \chi^{(1)} ) +2 \chi^{(2)} \, E_0 + 3 \chi^{(3)}\, E_0^2 + \cdots ] \, E_1 
 \end{equation}
is an effective dielectric function which will appear in the wave equation for the perturbation  $E_1$. 
If the classical field $E_0$ is a function of time, then $ \epsilon_{eff}$ will be also. If we quantize the perturbation,  in
general there will be mixing of the positive and negative frequency modes of the quantum field and
hence quantum creation of photons. 

\begin{figure}[htbp]
\includegraphics[scale=0.35]{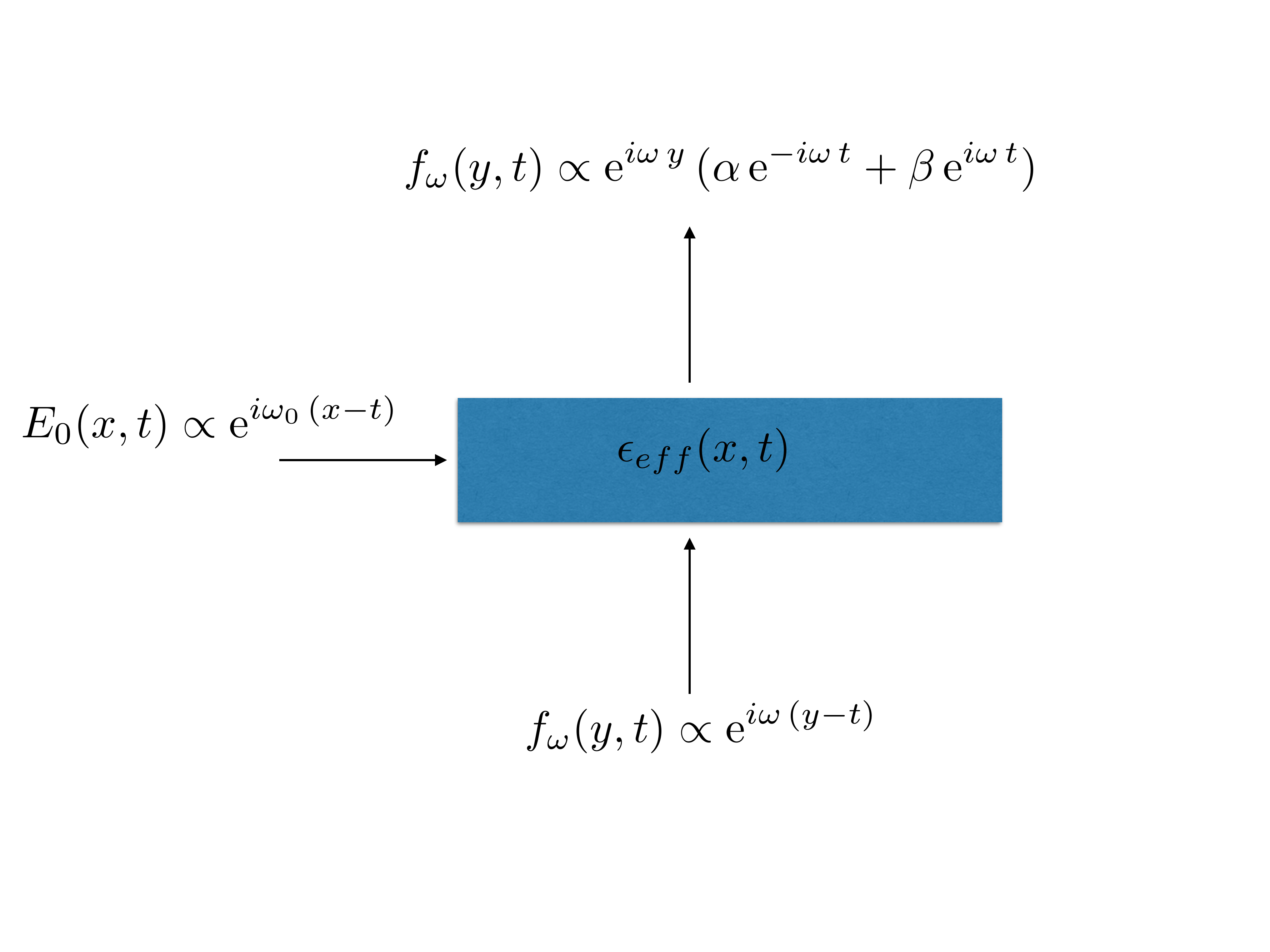}
\caption{The quantum creation of photons into a squeezed vacuum state in a nonlinear material is illustrated.
The shaded region is a slab of nonlinear material in which a classical field with angular frequency $\omega_0$
propagates. This creates a time dependent effective dielectric function, $\epsilon_{eff}$. A vacuum mode $f_\omega$
of the quantized electric field also propagates through the slab in a different direction. This mode enters the material
as a pure positive frequency mode, but exits as a superposition of positive and negative frequency parts, leading to
quantum photon creation. Both the classical field and the photons are assumed to be linearly polarized in the direction out
of the page. }
\label{fig:sq-light}
\end{figure}

Consider the geometry illustrated in Fig.~\ref{fig:sq-light}. Here the classical field $E_0$ is linearly polarized in the $z$-direction,
propagating in the $x$-direction, and has angular frequency $\omega_0$. It passes through a slab of nonlinear material and induces
a space and time-dependent effective dielectric function, $\epsilon_{eff}(x,t)$. A vacuum mode $f_\omega(y,t)$  of the quantized electromagnetic field,
with angular frequency $\omega$, enters the slab propagating in the  $y$-direction. Before entry, it is a pure positive frequency mode. Inside the
slab, it obeys a wave equation containing $\epsilon_{eff}(x,t)$, which leads to $f_\omega(y,t)$ becoming a superposition of positive and negative 
frequency components when it exits the slab. This is analogous to the effect of the expansion of the universe, discussed in Sect,~\ref{sec:particle-creation},
and leads to quantum creation of pairs of photons in a squeezed vacuum state. In both cases, particle number is not well-defined in the time-dependent 
region, and the energy needed to create particles comes from the time-dependent background.
The mean number of photons created into mode $f_\omega(y,t)$ will
be $|\beta|^2$, where $\beta$ is the amplitude of the negative frequency component of $f_\omega(y,t)$ in the final region. The angular frequency of 
the created photons depends both upon the frequency  $\omega_0$ of the classical field, and upon the order of the relevant nonlinear susceptibility.  
If second order nonlinearity,  described by $\chi^{(2)}$, dominates, then $\omega = \omega_0/2$. This was the case in the experiment reported
in Ref.~\cite{WKHW86}, so a pair of photons, each with energy  $\omega_0/2$, are created by a classical oscillating background with angular
frequency $\omega_0$. This is the exact analog of the creation of pairs of gravitons by an oscillating background described in Sec.~\ref{sec:late-time}.  
More general discussions of the effects of a time-dependent effective dielectric function have been given in Refs.~\cite{Dodonov93,Prain17}.

\subsection{Other Analog Models}

The various models discussed in the previous subsections can all be described as examples of the phenomenon of quantum vacuum friction,
of which there are other examples, such as atoms moving near a surface~\cite{Davies05}. The collapse or expansion of a Bose-Einstein
condensate bears some similarities to the cosmological expansion, and the excitations created in the condensate are analogous to
cosmological particle creation~\cite{Calzetta-Hu03,Prain10}. Several other analogies between superfluids and cosmology were discussed
by Volovik~\cite{Volovik01}. The creation of squeezed states of ions in a trap as an analog model was
discussed by Fey {\it et al}~\cite{Fey18}.   In another model, Walud {\it et al}~\cite {Walud17} have discussed the creation of low mass bosons
by an oscillating charged particle as a possible means for axion detection, as well as an analog of cosmological particle creation.

\section{Summary}
\label{sec:final}

This review has covered  the quantum creation of particles by the  expansion of the Universe using the formalism of quantum field theory in curved
spacetime. A quantum field theory requires both a set of local field operators, as well as a characterization of the set of quantum states. The latter
is especially subtle in an expanding universe, where no unique local definition of particles, and hence of the vacuum state exists. We have dealt with 
this ambiguity by use of quasilocal definitions in regions where the expansion rate is sufficiently small, and by asymptotic definitions when
available. However, it is this ambiguity which leads to the phenomenon of quantum particle creation. The definition of a particle in an initial region
can differ from that in a final region, so a quantum state without particles initially may later contain particles. This was discussed in Sec.~\ref{sec:Bogolubov}
using the formalism of Bogolubov transformations. The particle creation is also due to the non-conservation of field energy in a time-dependent
spacetime, but while the creation is occurring only approximate definitions of local particle number are meaningful. The inherent ambiguity of the
concept of particle number in a time dependent situation has observable consequences, as illustrated by the creation of squeezed states described
in Sec.~\ref{sec:squeeze}.
This review has concentrated on the creation of bosons, and especially scalar particles for
simplicity. Some exact results for particular model universes, as well as approximate results of more general spacetimes were reviewed, which 
allowed us to make some estimates of the particle creation rate. 

The quantum state of the created particles was treated in Sec.~\ref{sec:state-entropy}, where we found that these particles exhibit a high degree
of correlation and entanglement. We saw that the quantum state is a multi-mode squeezed vacuum state, which is a superposition of various numbers of
pairs of particles. Each pair contains particles with opposite momenta of equal magnitude. Interactions can lead to decoherence and entropy production.

The application of quantum particle creation to various cosmological models was discussed. In particular, quantum particle creation could play a role in the 
creation of matter in the universe after the end of an inflationary era, as discussed in Sec.~\ref{sec:inflation}.  There has been considerable work recently 
on models in which the dark matter arose from quantum particle creation.  Gravitons could be among the created particles, and a bath of relic gravitons
might leave a detectable imprint on the cosmological microwave background in the form of tensor perturbations. Cosmological anisotropy, inhomogeneity,
and rapid metric oscillations are other effects which could enhance particle creation.

The backreaction of the created particle on the background spacetime was discussed in Sec.~\ref{sec:backreact}. In a loose sense, the energy to create
the particle comes from the classical time-dependent gravitational field of the expanding universe. However, the energy of  gravitational field is difficult to
define clearly. A better description of the backreaction comes from the semiclassical theory of gravity, where the renormalized expectation value of the
quantum matter stress tensor acts as the source of a classical gravitational field.

Several analog models for cosmological particle creation were treated in Sec.~\ref{sec:analog}. These include photon creation by moving mirrors and other
time-dependent  media. Some of these models are accessible to experimental confirmation, which effectively provides support for the basic formalism
reviewed in this article. One example is the creation of squeezed states of light in nonlinear optical materials discussed in Sec.~\ref{sec:squeeze} which
is now routinely performed in the laboratory, and gives rise to squeezed vacuum states of the form expected in cosmological particle creation. This, and
several other topics treated in this review, are currently active areas of research. 

\begin{acknowledgments} 
The preparation of this review was supported in part  by the National Science Foundation under Grant PHY-1912545.
\end{acknowledgments}

\end{document}